\documentclass[journal=jacsat,manuscript=article]{achemso}

\usepackage{achemso} 
\usepackage{amssymb}
\usepackage{caption,float}
\usepackage{graphicx}
\usepackage{xcolor}
\usepackage[draft]{fixme}

\usepackage{subcaption}
\usepackage{mwe}
\usepackage{amsmath}
\usepackage{latexsym}
\usepackage{float}
\usepackage{enumerate}
\usepackage{amsfonts}
\usepackage{amssymb}
\usepackage{color}

\author{Peter Schurtenberger} \email{peter.schurtenberger@chem.lu.se}
\affiliation{Physical Chemistry, Department of Chemistry, Lund University, SE-221 00 Lund, Sweden}
\altaffiliation{LINXS Institute of Advanced Neutron and X-ray Science, Scheelevägen 19, SE-223 70 Lund, Sweden}

 \author{Marco Polimeni}
 \affiliation{Physical Chemistry, Department of Chemistry, Lund University, SE-221 00 Lund, Sweden}

\author{Sophia Marzouk}
\affiliation{Manchester Institute of Biotechnology, Department of Chemical Engineering, Faculty of Science and Engineering, The University of Manchester, Manchester M1 7DN, U.K.}

\author{Robin Curtis}
\affiliation{Manchester Institute of Biotechnology, Department of Chemical Engineering, Faculty of Science and Engineering, The University of Manchester, Manchester M1 7DN, U.K.}

 \author{Emanuela Zaccarelli} 
\affiliation{Institute for Complex Systems, National Research Council (ISC-CNR), Piazzale Aldo Moro 5, 00185 Rome, Italy}
\alsoaffiliation{Department of Physics, Sapienza University of Rome, Piazzale Aldo Moro 2, 00185 Rome, Italy}

 \author{Anna Stradner} 
\affiliation{Physical Chemistry, Department of Chemistry, Lund University, SE-221 00 Lund, Sweden}
\altaffiliation{LINXS Institute of Advanced Neutron and X-ray Science, Scheelevägen 19, SE-223 70 Lund, Sweden}

\title[An improved colloid model]
  {A soft penetrable sphere colloid model for the description of charge and excluded volume interactions in antibody solutions}


\begin{document}




\begin{abstract}

Colloid models have frequently been used to successfully describe the influence of protein-protein interactions on antibody solution properties, but they suffer from inherent problems due to the anisotropic shape of the particles. The net charge required to describe electrostatic interactions is an effective quantity that cannot directly be obtained from the known molecular structure of an antibody, and the solution structure caused by excluded volume interactions is strongly overestimated at high concentrations due to the assumption of hard sphere interactions. As a result, these models have descriptive rather than predictive power. Here we present an improved, soft penetrable sphere model based on analogies to soft colloids and star polyelectrolytes that take into account the Y-shaped antibody form and the corresponding charge and ion distribution. The model not only correctly describes the concentration and ionic strength dependence of thermodynamic and collective dynamics quantities such as the osmotic compressibility and the apparent hydrodynamic radius, but also reproduces the center-of-mass static structure factor obtained in computer simulations using a weakly coarse-grained model, in which the antibody is described at an amino acid level. 
We demonstrate that this soft penetrable sphere model quantitatively reproduces experimental data from static and dynamic light scattering at low and high ionic strength for two well-characterized monoclonal antibodies (mAbs) using the net charges and the overall mAb dimensions directly obtained from their molecular structure.   
\end{abstract}

\section{Introduction}
The use of colloid models to describe protein-protein interactions and their effect on measurable structural and dynamic quantities such as the isothermal osmotic compressibility $\kappa_T$, the static structure factor $S(q)$, the apparent hydrodynamic radius $R_{h,app}$ or the relative viscosity $\eta_r$ has a long tradition in protein science \cite{muschol1997liquid,woldeyes2017predicting,stradner2020potential}. In such an approach, the molecular structure of the protein is replaced by a spherical colloid interacting via an isotropic effective pair potential $U_{eff}(r)$ (or potential of mean force, PMF) that typically consists of contributions from hard core or excluded volume, charge, and short-range attractive interactions. Here, the charge interactions are usually expressed via a screened Coulomb or Yukawa potential, and the short-range attraction contains contributions from van der Waals and hydrophobic interactions. For compact globular proteins such as bovine serum albumin\cite{Neal1999}, lysozyme\cite{Madani2025}, or the eye lens crystallin family\cite{Foffi2014,Bucciarelli2015} this has indeed been highly successful. The parameters used to build $U_{eff}(r)$ for these systems have been in good agreement with predictions based on the actual molecular structure and allowed the reproducibility of key solution properties, as well as the phase behavior characterized by the phase boundaries for liquid-liquid phase separation and crystallization 
However, there were also reports that for some quantities that are particularly sensitive to the details of protein-protein interactions, these simple colloid models failed \cite{stradner2020potential, Bucciarelli2016, Bergman2025}. This failure of the colloid analogy was attributed to anisotropic contributions to protein-protein interactions caused by attractive patches~\cite{fusco2016soft,mcmanus2016physics}, an assumption that was supported by the results of theory \cite{gogelein2008simple}and simulations\cite{Bucciarelli2016, Myung2018} using more refined patchy models. 

Colloid models have also been used frequently to describe protein-protein interactions in solutions of monoclonal antibodies (mAbs)\cite{roberts2014role,calero2018predicting,sibanda2023relationship,chowdhury2023characterizing}, albeit with mixed success. Although these models were found to successfully reproduce key solution parameters such as $\kappa_T$, $R_{h,app}$, second virial coefficient $B_2$, or diffusion interaction parameter $k_D$, the effective pair potential required to describe the experimental results was not in agreement with estimates based on the molecular structure of the mAbs investigated. In particular, the effective charge $Z_{eff}$ required to reproduce the experimental data was generally significantly below the net charge predicted from the known molecular structure~\cite{Roberts2014}. Initially, these discrepancies were interpreted primarily as the result of ion adsorption~\cite{Roberts2014}. However, recent investigations pointed out that they are also caused by the intrinsic inadequacy of the underlying model used for electrostatic contributions to protein-protein interactions~\cite{gulotta2024,Polimeni2024}. In a systematic coarse-graining study using Y-shaped bead models the authors showed that the effective charge required to reproduce the measured structure factors $S(q)$ depends on the level of coarse-graining. In particular, $Z_{eff}$ was found to systematically increase from the value obtained for the one bead colloid model, reaching the net charge predicted from the molecular structure for a weakly coarse-grained structure where each amino acid is represented by a bead at a high enough number of beads\cite{Polimeni2024}.

\begin{figure}[!h]
\includegraphics[width=0.9\linewidth]{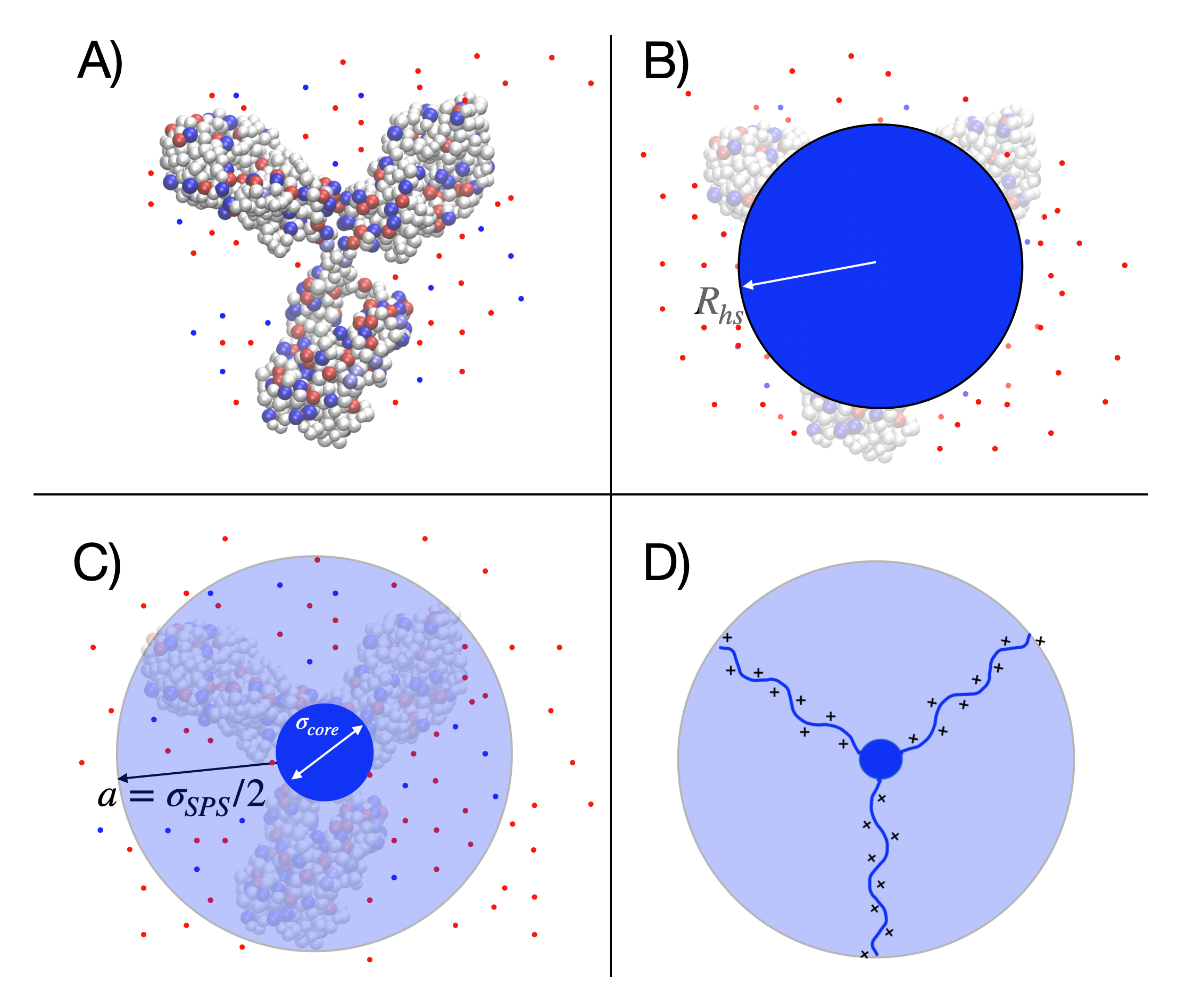}
\caption{Differently coarse-grained representations of the Y-shaped antibody, consisting of two arms (Fab domains) and one leg (Fc domain). A: Image of the charge distribution. Here, the antibody is slightly coarse-grained, at the amino acid (aa) level.  Thus, each bead represents an amino acid with coloured beads referring to charged amino acids, with blue corresponding to positive (+1e) and red to negative charges (-1e). Also shown are the positive and negative counterions as small blue and red spheres, respectively. B: Schematic representation of an effective hard sphere model of the monoclonal antibody, together with its aa model. C: Soft penetrable sphere model. The particles have a hard core with diameter $\sigma_{core}$, and a soft shell with diameter $\sigma_{SPS}$ given by the overall size of the mAb, where other particles and small ions can penetrate. D: Star polyelectrolyte model used for the calculation of the electrostatic contribution to the potential of mean force between mAbs.
}
\label{CollModel}
\end{figure}

The reason for the deficiency of the colloid model in accurately describing electrostatic interactions based on the correct net antibody charge is illustrated in Fig.\ref{CollModel}.
In the colloid model, all charges are distributed on the surface of a spherical particle with a radius $R_{hs}$, and the electrostatic contribution to the interaction potential is described by a screened Coulomb or Yukawa potential, Fig. \ref{CollModel} B) . In contrast, for a Y-shaped object, such as a mAb, antibody charges and counterions will also be present within the enclosing sphere around the mAb  (Fig. \ref{CollModel} A)). In calculating the electric field or the resulting interaction potential between two particles at distances larger than the diameter of a mAb, one thus also needs to consider the explicit radial macroion charge density and the resulting counterion distribution within this enclosing sphere. 
As we will show in detail below, while the long-range tail of the interaction potential is still described by a Yukawa potential, this results in a prefactor that can be significantly smaller for the Y-shaped object than for the classical DLVO-type potential of a spherical colloid, resulting in a smaller interaction potential for distances larger than the particle diameter for the same effective charge. 

When looking at Fig. \ref{CollModel} A), the charge and counterion distribution bears some similarity to polyelectrolyte star polymers, for which Denton has derived the corresponding potential as a function of charge and ionic strength \cite{Denton2003}. We therefore explore the application of an improved colloid model for the description of protein-protein interactions in antibody solutions inspired by soft colloids such as polyelectrolyte stars or microgels. We first use computer simulations using an amino acid level (aa) coarse-grained model as a reference system to obtain an effective potential of mean force for a specific monoclonal antibody. We then construct an effective potential based on a model of soft penetrable spheres (SPS) with a hard core of diameter $\sigma_{core}$, and a soft shell with diameter $\sigma_{SPS}$ given by the overall size of the mAb, where other particles and small ions can penetrate, as shown in Fig. \ref{CollModel} C). Here we assume that the outer shell has a radial charge density $\rho \sim r^{-2}$ analogous to that of a 3-arm star polyelectrolyte. We compare the resulting interaction potential with the PMF from the computer simulations for the reference mAb model. This soft sphere potential is then used to calculate the concentration and ionic strength dependence of key solution properties such as $\kappa_T$, $S(q)$, $R_{h,app}$, $B_2$, and $k_D$, which are subsequently compared with the corresponding values obtained from computer simulations and experiments.

\section{Materials and methods}

We compare our theoretical results against a detailed characterization of a mAb called Actemra (or Tocilizumab), which has previously been published by some of us \cite{gulotta2024}. Actemra, which we subsequently label mAb-1, is an IgG1 that is an anti-IL-6 receptor with a pI of 9.18. Measurements were made with two buffer solutions at different ionic strengths, \emph{i.e.}, 7 and 57 mM at pH = 6. 

Additional experiments and simulations were also performed for the NIST mAb as a reference model for a highly stable mAb, which we subsequently label mAb-2. The NISTmAb Primary Sample (PS 8670) was obtained from NIST (Gaithersburg, MD, USA). NISTmAb was received at 100 mg/ml. All experiments were conducted in D$_2$O-based buffer. Protein samples were buffer-exchanged into D$_2$O containing sodium acetate (pH~5) at an ionic strength of 10~mM. The mAb formulation (100~g~L$^{-1}$) was first diluted tenfold with D$_2$O buffer and then subjected to six successive centrifugal ultrafiltration cycles (Amicon Ultra, 30~kDa MWCO), each corresponding to an approximately tenfold concentration step. Buffer exchange was deemed complete once the solution pH stabilised.  The final protein solution was filtered (0.10~$\mu$m), and the protein concentration was determined by UV absorbance spectroscopy (NanoDrop One, Thermo Scientific) using an extinction coefficient equal to 1.42~L/(g-cm). The stock protein solution was adjusted to 24~g~L$^{-1}$ with D$_2$O buffer. A 1~M NaCl stock solution was prepared in the same buffer.

Dynamic light scattering (DLS) measurements for mAb-2 were performqed using a \textit{DynaPro Plate Reader} (Wyatt Technology, USA) equipped with an 830~nm laser and a single-photon counting detector positioned at 158°. Samples spanning a range of protein concentrations at fixed NaCl concentration were prepared by diluting a stock protein solution with protein-free buffer at the same salt concentration. Triplicate samples were loaded into a 384-well microplate and measured under temperature-controlled conditions. Measurements consisted of ten 5~s acquisitions per well, with an intensity autocorrelation function calculated for each acquisition. The resulting correlation functions were analysed using the cumulants method implemented in \textit{DYNAMICS} software (Wyatt Technology) to obtain the translational diffusion coefficient $D$ and the polydispersity index $P_{\rm d}$. Measured polydispersities were consistently below 8\%, indicating a monodisperse protein solution.

The mutual diffusion coefficient $D$ was determined as a function of protein concentration and analysed according to
\begin{equation}
D = D_0 \left(1 + k_D c \right),
\end{equation}
where $D_0$ is the diffusion coefficient at infinite dilution, $c$ is the protein concentration, and $k_D$ is the diffusion interaction parameter. Values of $k_D$ were obtained from the slope of $D/D_0$ versus concentration.

Static light scattering (SLS) measurements for mAb-2 were performed using a \textit{miniDAWN TREOS} detector (Wyatt Technology, USA) equipped with a flow cell and operated at a wavelength of 690~nm. Scattering intensities from the 90° detector were used for analysis. Sample preparation and delivery were automated using a \textit{Calypso II} three-pump syringe system (Wyatt Technology, USA). Protein, buffer, and salt stock solutions were filtered inline (0.1~$\mu$m) and combined via a static mixer immediately prior to entering the flow cell. Protein concentration was varied stepwise from 5 to 0.5~mg~mL$^{-1}$ to obtain concentration-dependent scattering intensities, while salt concentrations were adjusted by varying the flow rate of the stock salt stream.

The static structure factor in the zero–scattering vector limit, $S(0)$, was determined from
\begin{equation}
\frac{K^{*} c}{\overline{R_\theta}} = \frac{1}{M S(0)}
\end{equation}
where $c$ is the protein concentration, $\overline{R_\theta}$ is the excess Rayleigh ratio relative to solvent, and $M$ is the protein molecular weight. The optical constant $K^{*}$ is proportional to the square of the protein refractive index increment, for which a value of 0.185~mL~g$^{-1}$ was used. Molecular weights obtained from extrapolation agreed with the sequence value within $\pm$3~kDa.

\subsection{Coarse-graining of mAb structures at the amino acid level }
Starting from the atomistic PDB structures of mAb-1 and mAb-2, we built their coarse-grained versions at the amino acid (aa) level using the same protocol as in Ref.~\cite{Mahapatra2022} by replacing each amino acid with a spherical bead of diameter $\sigma_{i} = (6M_W/\pi \rho)^{1/3}$, where $M_W$ is the amino acid molecular weight (in g mol$^{-1}$), and $\rho = 1$ (in g mol$^{-1}$ Å$^{-3}$ ) is an average amino acid density \cite{Kaieda2014WeakSO}. For mAb-1, the construction of the aa-level coarse-grained model is described in detail in ref.  \cite{Polimeni2024}. For mAb-2, the initial structure was retrieved from https://www.nist.gov/programs-projects/nist-monoclonal-antibody-reference-material-8671, and the coarse-graining was subsequently made following the same protocol as for mAb-1.

\subsection{Constant-pH one-protein simulations}
With the mAbs coarse-grained at the amino acid level, we perform Monte Carlo (MC) simulations using Faunus \cite{Stenqvist2013FaunusA} to calculate their charge distributions and net charge at a given pH and ionic strength. 
For each mAb,  we have carried out constant-pH MC simulations with a single rigid mAb, and titration moves only \cite{Johnson1994ReactiveCM}, which allow the amino acid charges to fluctuate and to reach an equilibrium distribution. The titration move propagates back and forward the reaction  $AH \rightleftarrows A+H$ using a reactive MC scheme \cite{Johnson1994ReactiveCM}, where $AH$ is the protonated form of the amino acid, while $A+H$ is its dissociated form. The contribution of the titration move to the energy change in the MC scheme is given by:
\begin{eqnarray}
    \beta \Delta U = &-& \sum_{i} \ln \left( \frac{N_{i}!}{(N_{i} +\nu)!}V^{\nu_{i}} \right) -  \ln\prod_i a_{i}^{\upsilon_{i}},
\label{eq:titration}    
\end{eqnarray}  
\noindent where $\beta = 1/k_BT$ is the inverse thermal energy, $N_i$ is the number of charged amino acids of species i, $\nu_i$ is the stoichiometric coefficient (positive for the products and negative for the reagents), $V$ is the volume of the system, and $a_{i}$ is the activity of the amino acids of type $i$ (see Faunus documentation: https://faunus.readthedocs.io/). The MC move shifts the equilibrium reaction based on the solution conditions (such as pH and ionic strength) and the equilibrium constants associated with each reaction, $K_{\mathrm{a},i}$.  Values for $K_{\mathrm{a},i}$ were taken from the literature
\cite{Thurlkill2006}. For each simulation, we performed $10^4$ MC sweeps where, on each sweep, $N = 10 $ titration moves were attempted. The charge distributions obtained from these calculations are represented in SI Figs. S1 and S2 for mAb-1 and mAb-2, respectively.

\subsection{Two-protein simulations - potential of mean force}

Using the mAbs' amino acid representation, we performed two-protein MC simulations to sample the mAbs' potential of mean force (PMF). The simulation setup is shown in SI Fig. S3. The center of mass of a first mAb (left) is fixed while a second (right) is initially placed 150 \r{A} away along the Z-axis. The first mAb can only rotate around its center of mass, while the second could also rigidly translate along the Z-axis.

MAbs interact with each other through non-bonded bead-bead potentials. We used either a pure hard-sphere potential, $U^{HS}_{ij}$,
\begin{equation}
\begin{centering}
    U^{HS}_{ij}= 
\begin{cases}
    \infty,& r_{ij} \leq \sigma_{ij} \\
    0,              & r_{ij} > \sigma_{ij}
\end{cases}
\label{eq:harmonic}
\end{centering}
\end{equation}
or a combination of a Yukawa potential that describes screened electrostatic interactions and a Lennard-Jones potential, which accounts for short-range attractive interactions and steric repulsions, 
\begin{equation}
\begin{centering}
\beta U_{ij} = \frac{\lambda_Bz_iz_j}{r_{ij}} e^{-\kappa r_{ij}}
+ 4 \varepsilon_{ij} \left[ \left(\frac{\sigma_{ij}}{r_{ij}}\right)^{12}- \left(\frac{\sigma_{ij}}{r_{ij}}\right)^{6} \right].
\label{eq:ham}
\end{centering}
\end{equation}

\noindent Here, $\lambda_B=\beta e^2/4\pi\epsilon_0\epsilon_r$ is the Bjerrum length, with $\epsilon_0$ the vacuum permittivity, $\epsilon_r=80$ the relative permittivity of water (being the solvent treated implicitly), $e$ the (positive) electron unit charge, and $\beta=1/k_BT$ the inverse thermal energy, with $k_B$ Boltzmann's constant and $T=300$~K the temperature. In addition,  $z_i$ and $z_j$ are the charges of the $i^{th}$ and $j^{th}$ bead,  $r_{ij}$ the distance between them and  $\sigma_{ij} = (\sigma_{i} + \sigma_{j})/2$. The energy depth of the Lennard-Jones interactions, $\beta \varepsilon_{ij} = 0.075$, was fixed to reproduce the experimental small-angle X-ray scattering (SAXS) curves at high salt concentration.  The values for $\sigma_{ij}$ are reported in Table S1, Supporting Information. Finally, $\kappa$ corresponds to the inverse of the Debye length, $\lambda_{D}$, which accounts for salt screening effects (with the salt treated implicitly), and is calculated through the relation:
\begin{equation}
    \kappa^2 = 1/\lambda_{D}^{2} =  4 \pi \lambda_{B} \left[ Z_{mAb}\rho_{mAb} + 2\rho_{salt} + 2\rho_{buffer}\right],
\label{eq:kappa}
\end{equation}
where $\rho_{mAb}$ is the mAb number density (which is zero for two-mAb simulations as we are assuming infinite dilution), Z\textsubscript{mAb} is the mAb net charge, and $\rho_{salt}$ and $\rho_{buffer}$ are, respectively, the number densities of salt and of the dissociated buffer ions. Using the simulation setup shown in Fig. S3 in SI, we sample the radial distribution function of the mAbs, $g(r)$, and, consequently, the potential of mean force (PMF), $U(r)$, given by 
\begin{equation}
U(r) = -k_B T \ln g(r).
\end{equation}

\subsection{Many-protein MC simulations}
Using the mAb representations at the amino acid level, we further perform many-protein MC simulations. For each condition, $N_{p}$ = 20 rigid translating and rotating mAbs are inserted in a cubic box of volume $V = N_{p}M_{w}/(c_{p}N_{a}10^{-27})$, where $M_{w}$ is the mAb molecular weight in kDa, $N_{a}$ is Avogadro's number, and $c_{p}$ is the mAb concentration (see SI Fig. S4). mAbs interact again either with a hard-sphere potential (Eq.~\ref{eq:harmonic}) or a screened Coulomb plus a Lennard-Jones potential (Eq.~\ref{eq:ham}). 
For each of the solution conditions, we perform five replicas, 
each is carried out for $10^4$ MC sweeps, where, on each sweep, each mAb is attempted to be translated and rotated. 

The purpose of the many-body simulations was to determine the center of mass structure factor $S^{cm}(q)$ and the effective structure factor measured in scattering experiments, $S^{eff}(q)$, of the mAb solution under different conditions. $S^{cm}(q)$ was calculated as:
\begin{equation}
S^{cm}(q) = \frac{1}{N} \left\langle \sum_{i,j}^{1,N}  
    \exp^{i\vec{q}\cdot(\vec{r}^{CM}_i-\vec{r}^{CM}_j)}
    \right\rangle =  \frac{1}{N}
\left\langle \left( \sum_{i}^{N}  \sin({\textbf{q}\textbf{r}^{CM}_{i}})\right)^2 + \left( \sum_{i}^{N}  \cos({\textbf{q}\textbf{r}_{j}^{CM}})\right)^2 \right\rangle 
\label{eq:Sq-cm}
\end{equation}
where $N$ is the number of mAbs and $r^{CM}_{i}$ and $r^{CM}_{j}$ identify the position of the \textit{$i^{th}$} and \textit{$j^{th}$} mAb center of mass. Similarly, $S_{eff}(q)$ was calculated as: 
\begin{equation}
S_{eff}(q) = \frac{1}{F(q)} \frac{1}{N_{tot}}
\left\langle \left( \sum_{i}^{N_{tot}}  \sin({\textbf{q}\textbf{r}_{i}})\right)^2 + \left( \sum_{i}^{N_{tot}}  \cos({\textbf{q}\textbf{r}_{j}})\right)^2 \right\rangle 
\label{eq:Sq-bead}
\end{equation}
where $N_{tot}$ is the product of the number of mAbs and the number of beads present on each mAb, while $r_{i}$ and $r_{j}$ now identify the position of the \textit{$i^{th}$} and \textit\textit{$j^{th}$} bead, and $F(q)$ is the form factor of the single molecule. In both Eqns.~\ref{eq:Sq-cm} and ~\ref{eq:Sq-bead}, the average is performed over all directions obtained by permuting the crystallographic index [100], [110], [111] of the cubic box to define the scattering vector $\textbf{q}=2\pi p/L(h,k,l)$, where $p=1,2,…,p_{max}$, and $p_{max}=25$.

\section{Results and discussion}

We have previously shown that the aa-level coarse-grained representation of mAb-1 is capable of reproducing the experimental static structure factors $S(q)$ determined with SAXS quantitatively over the full range of concentrations and ionic strengths investigated \cite{Polimeni2024}. We therefore use this model as a reference state and determine the corresponding PMF and the resulting structural properties. The PMFs are sampled from MC computer simulations as described in Materials and Methods. Computer simulations allow us to selectively investigate the contributions from different interactions, and we therefore determine these quantities separately for hard sphere interactions only, and for the full interaction potential between different beads. 

\subsection{Hard sphere interactions}

\subsubsection{Potential of mean force}

\begin{figure}[!h]
\includegraphics[width=0.9\linewidth]{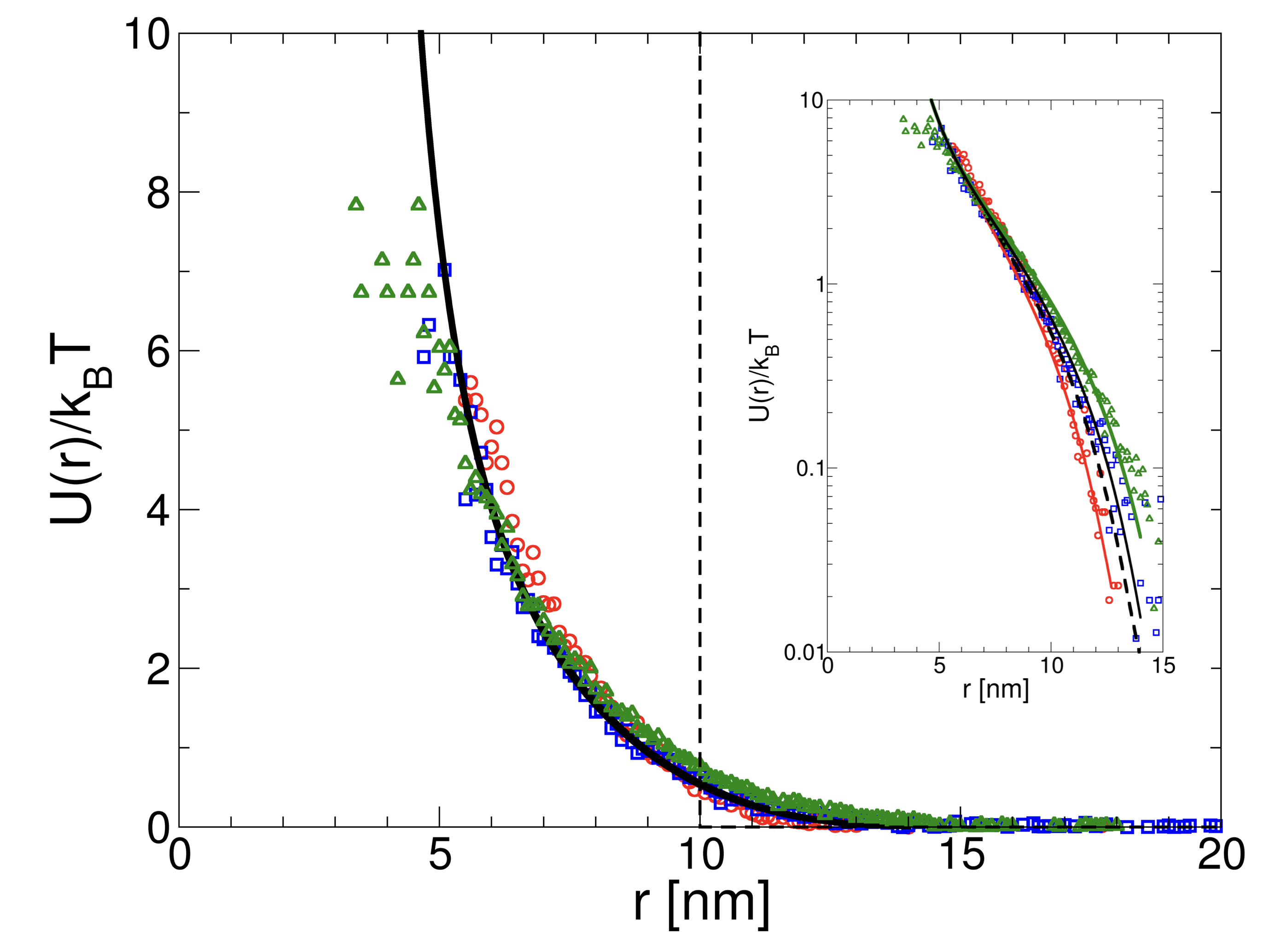}
\caption{Potential of mean force (PMF) as a function of the center-center distance $r$ obtained from MC simulations using an aa-level coarse grained model for three different mAb structures where beads interact with a hard sphere potential only: mAb-1 structure at 7 mM ionic strength (blue squares), mAb-1 structure at 57 mM ionic strength (red circles), and mAb-2 structure (green triangles). Also shown are the PMF for the SPS model with an interaction potential given by Eq.~\ref{eq:SPS-ev-pot} as the solid black line (generic parameter values as given in table \ref{tab:parameter-hs}), and that of a hard sphere colloid with hard sphere diameter $\sigma_{hs} = 10$ nm as the black dashed line. The inset shows the data for all three mAb structures in a log-lin representation together with the corresponding SPS potentials using the individually optimized parameters given in Table~\ref{tab:parameter-hs}. The data for the generic model is shown as the black dashed line.}
\label{fig:PMFs}
\end{figure}

We first focus on excluded volume or hard sphere interactions only, given by Eq.~\ref{eq:harmonic}.
We have calculated the corresponding PMFs for hard sphere interactions using a reference aa-coarse-grained model for three different mAb conformations: Two for mAb-1, where the atomistic simulations yielded different representative conformations for the two ionic strengths (7 mM and 57 mM), as described in detail in ref. \cite{Polimeni2024}), and one for mAb-2. The corresponding results from the two-protein MC simulations are shown in Fig. \ref{fig:PMFs}.

The PMFs for the aa coarse-grained models almost completely overlap for all three structures, with only small variations related to the slightly different overall dimensions and local structure of the different mAb conformations. The PMF due to excluded volume effects between mAbs appears much softer than the commonly used colloidal hard sphere potential (Eq.~\ref{eq:harmonic}), which is also shown in Fig. \ref{fig:PMFs} for a colloid with $\sigma_{hs} = 10$ nm. This value is chosen to be equal to $2R_g$, where $R_g$ is the radius of gyration of the mAb, since it previously allowed us to describe experimental data for mAb-1 \cite{gulotta2024}. 

The classical hard sphere colloid model (black dashed line in Fig. \ref{fig:PMFs}) clearly fails to describe the simulated PMF, and we need another approach to account for excluded volume effects in mAb solutions. We thus construct a purely phenomenological expression for excluded volume interactions between mAbs that reproduces the data in Fig. \ref{fig:PMFs}, using other soft colloid systems such as star polymers and microgels as inspiration\cite{Likos1998,Bergman2018}. We model the mAb as a soft penetrable particle with an overall diameter $\sigma_{SPS}$, which is given by the diameter of the circumscribing sphere, and a hard impenetrable core of diameter $\sigma_{core}$. We then consider two contributions, one from the outer penetrable region, and one from the core. The first is written as a modified Hertzian of the form $U_{ev,outer}(r) \sim (1-r/\sigma_{SPS})^{a_1}$ for $r < \sigma_{SPS}$, while the inner contribution is approximated as a power law of the form $U_{ev,inner}(r) \sim (\frac{\sigma_{core}}{r})^{a_2}$. The resulting functional form is,

\begin{equation}
    \label{eq:SPS-ev-pot}
\begin{aligned}
  \beta U_{SPS,ev}(r) = \ & \ \infty, &(r < \sigma_{\textrm{core}})\\
=\ & \  A_1\left(1 - \frac{r}{\sigma_{SPS}}\right)^{a_1} + A_2\left(\frac{\sigma_{core}}{r}\right)^{a_2} &(\sigma_{\textrm{core}}<r<\sigma_{SPS})\\
=\ & \ 0,  &(r> \sigma_{SPS})
\end{aligned}
\end{equation}

\noindent A comparison between the simulated PMFs and those obtained with this phenomenological model (using the parameters given in Table~\ref{tab:parameter-hs}) shows a very good agreement.

\begin{table}[h!]
\centering
\begin{tabular}{| c | c | c | c | c |c | c | c |} 
 \hline
   Sample & $A_1$  & $a_1$ & $\sigma_{SPS}$ [nm]  & $A_2$  & $a_2$ & $\sigma_{core}$ [nm] & $B_2^*$   \\ [0.5ex]  
  \hline
  mAb-1, 7 mM & 11.5 & 2.7 & 15  & 8.0 & 6.0 & 4.4 & 1.08    \\
   \hline
   mAb-1, 57 mM & 13.3 & 2.7 & 13.6  & 8.0 & 6.0 & 4.4 & 0.89       \\
   \hline
   mAb-2 & 10.7 & 2.7 & 15.9  & 8.0 & 6.0 & 4.4 & 1.23      \\
    \hline
   generic & 12.3 & 2.8 & 14.6  & 8.0 & 6.0 & 4.4 & 0.99    \\
          [1ex] 
 \hline
\end{tabular}
\caption{Parameters for the calculation of $U_{SPS,ev}$ for the SPS model using Eq.~\ref{eq:SPS-ev-pot} for the different mAb samples and solvent conditions. Also shown is the resulting normalised second virial coefficient $B_2^*$, where the reference hard sphere system is a hard sphere with diameter $\sigma_{hs} = 10$ nm. Also shown are the values for an average or generic excluded volume contribution for the SPS model that is later used for all conformations and mAbs.}
\label{tab:parameter-hs}
\end{table}

As shown in the inset of Fig. \ref{fig:PMFs}, using a log-lin representation, we observe some minor differences at large $r$-values that reflect the small differences in the overall dimensions of the different mAb conformations. These dependencies of the excluded volume interactions on the overall mAb diameter can be incorporated in the SPS model by rescaling $\sigma_{SPS}$ with the diameter of the circumscribing sphere and adapting $A_1$ accordingly, as shown in the inset of Fig. \ref{fig:PMFs}. However, given the small differences in $U_{ev}/k_BT$, we next investigate their importance for key solution properties.

\subsubsection{Osmotic compressibility and second virial coefficient}

We first calculate the second virial coefficient $B_2$ or the normalized second virial coefficient $B_2^* = B_2/B_2^{hs}$ as an integral description of the PMF. $B_2$ is given by \cite{McQuarrie2000}

 \begin{equation}
	\label{B2}
	B_2 = 2 \pi  \int_{0}^{\infty}  \bigg(1 - e^{-\beta U(r)}\bigg) r^2 \,dr\,
\end{equation}

\noindent and $B_2^{hs} = 4(\pi \sigma_{hs}^3/6)$ is the second virial coefficient of the corresponding pure hard sphere reference system. Here we use a value of $\sigma_{hs} = 10$ nm, in line with the previous work on mAb-1 \cite{gulotta2024, Polimeni2024}. The $B_2^*$ values calculated for the different PMFs using Eq.~\ref{eq:SPS-ev-pot} are also summarised in Table~\ref{tab:parameter-hs}. We indeed observe small variations in the resulting $B_2^*$ values. This is not surprising given the fact that $B_2$ corresponds to a volume integral (Eq.~\ref{B2}), which heavily weighs the weak contributions seen at large $r$-values. However, it is important to realise that even a difference of 23\% in $B_2^*$ corresponds to a difference of only 7\% in the corresponding effective hard sphere diameter $\sigma_{hs}$. Given the uncertainty in experimentally determining $R_g$ and the fact that we represent each mAb by a single average conformation that does not include intrinsic flexibility and conformational dynamics, we believe that it is not justified to individually optimize the parameters in Eq.~\ref{eq:SPS-ev-pot} for every mAb and different solution conditions, but instead use model parameters that are averaged over all calculated conformations and PMFs. Overall the different PMF curves are indeed well represented by the parameters of an average or `generic' model, also given in Table~\ref{tab:parameter-hs}, for which the resulting $B_2^*$ is almost one. This indicates that a simple hard sphere colloid model with $\sigma_{hs} =2R_g$ reproduces the excluded volume contributions to the second virial coefficient quite well, in agreement with previous observations \cite{gulotta2024}.

\begin{figure}[h]
\centerline{\includegraphics[width=0.9\linewidth]{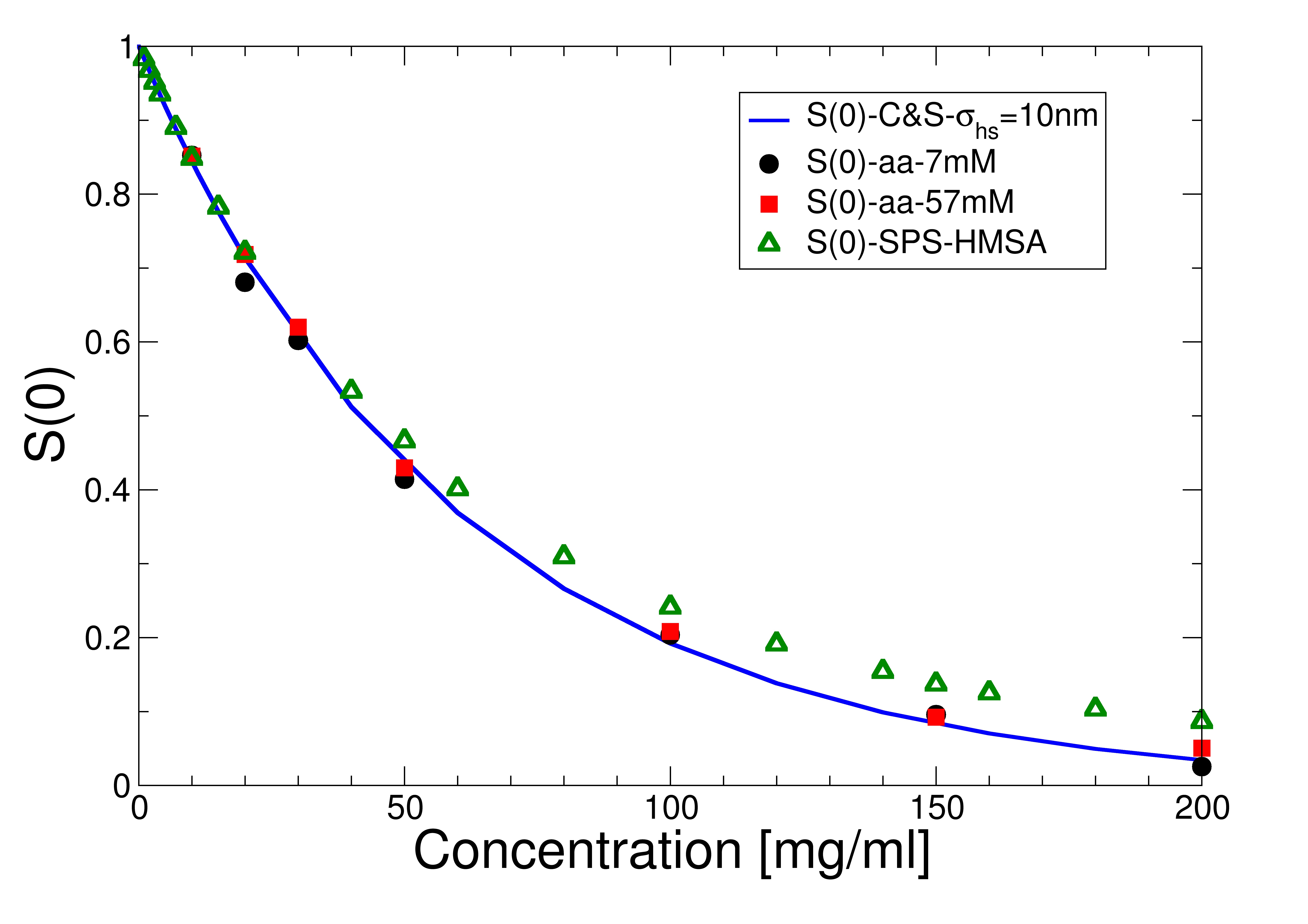}}
\caption{$S(0)$ vs. \emph{c} for different models using excluded volume interactions only. Shown are the theoretical results for hard spheres based on the approximation by Carnahan and Starling (eqn. \ref{eq:SCS}) and $\sigma_{hs} = 10$ nm as the blue line, and for the SPS model using the hybridized-mean spherical approximation (HMSA) closure relation and the PMF from the generic model parameters given in table \ref{tab:parameter-hs}. Also shown are the results from computer simulations for mAb-1 conformations corresponding to 7 mM and 57 mM ionic strength, respectively. The arrow indicates the concentration that corresponds to an effective volume fraction of $\phi_{eff} = 0.63$ for random close packing of the penetrable spheres with effective diameter $\sigma_{SPS}$.}
\label{fig:S0-hs}
\end{figure}

Next, we compare the predictions of the different models for the osmotic compressibility $\kappa_T$ expressed by the asymptotic low-$q$ value of the static structure factor $S(0)$. For the hard sphere colloid model, we use the Carnahan and Starling (CS) approximation for the hard sphere $S^{CS}(0)$ \cite{Carnahan1969}: 
\begin{equation}
	\label{eq:SCS}
	S^{CS}(0) = \frac{(1-\phi_{hs})^4}{( 1+2\phi_{hs})^2 + \phi_{hs}^3(\phi_{hs}-4)},
\end{equation}

\noindent where the hard sphere volume fraction $\phi_{hs}$ is given by 
\begin{equation}
	\label{eq:effhsvolume}
	\phi_{hs} = \rho \frac{\pi \sigma^3_{hs}}{6}
\end{equation}

\noindent with $\rho$ is the number density of particles. For the SPS model we calculate $S^{HMSA}(0)$ based on liquid state theory~\cite{Goodstein1975}, i.e. using integral equations with the hybridized-mean spherical approximation (HMSA) closure relation, that was first proposed by Zerah and Hansen~\cite{Zerah1986}, which should be particularly suitable for soft potentials with additional attractive contributions that we will have to consider later. Finally, we compare these results with the values obtained from the simulations with the aa-level coarse-grained mAb model as described in Materials and Methods. A comparison between the different models is shown in Fig. \ref{fig:S0-hs}.

Figure \ref{fig:S0-hs} confirms that for excluded volume interactions, the compressibility calculated for the hard sphere and the soft penetrable sphere models overlap at low concentrations. Moreover, both models reproduce the values obtained from the computer simulations using the aa bead model. At high enough concentrations, $c \gtrsim 50$ mg/ml, the values for the hard sphere colloid and the SPS models start to deviate. The hard sphere $S^{CS}(0)$ agrees well with the computer simulation results, while the SPS model appears to underestimate excluded volume effects at high concentrations. 

The disagreement between simulations and the SPS model at high concentrations is not surprising. The model is based on spherical particles interacting via an isotropic potential. While this assumption appears justified at low concentrations, where the Y-shaped mAbs can freely rotate and primarily experience the soft tail of the PMF at larger distances, this is no longer the case at higher concentrations. Here, the mAbs start to overlap and are no longer able to freely rotate. At short distances $r < \sigma_{SPS}$, the excluded volume interactions between mAbs will then strongly depend on the orientation of the mAbs and thus become anisotropic. It is interesting to calculate the overlap concentration for the SPS model. This should happen for $\phi_{eff,SPS} \approx 0.63$, corresponding to random close packing of spheres with diameter $\sigma_{SPS}$. The overlap concentration should thus correspond to $c \approx 95$ mg/ml, also shown in Fig. \ref{fig:S0-hs}.

\subsubsection{Static structure factor}

\begin{figure}[h!]
\includegraphics[width=0.6\linewidth]{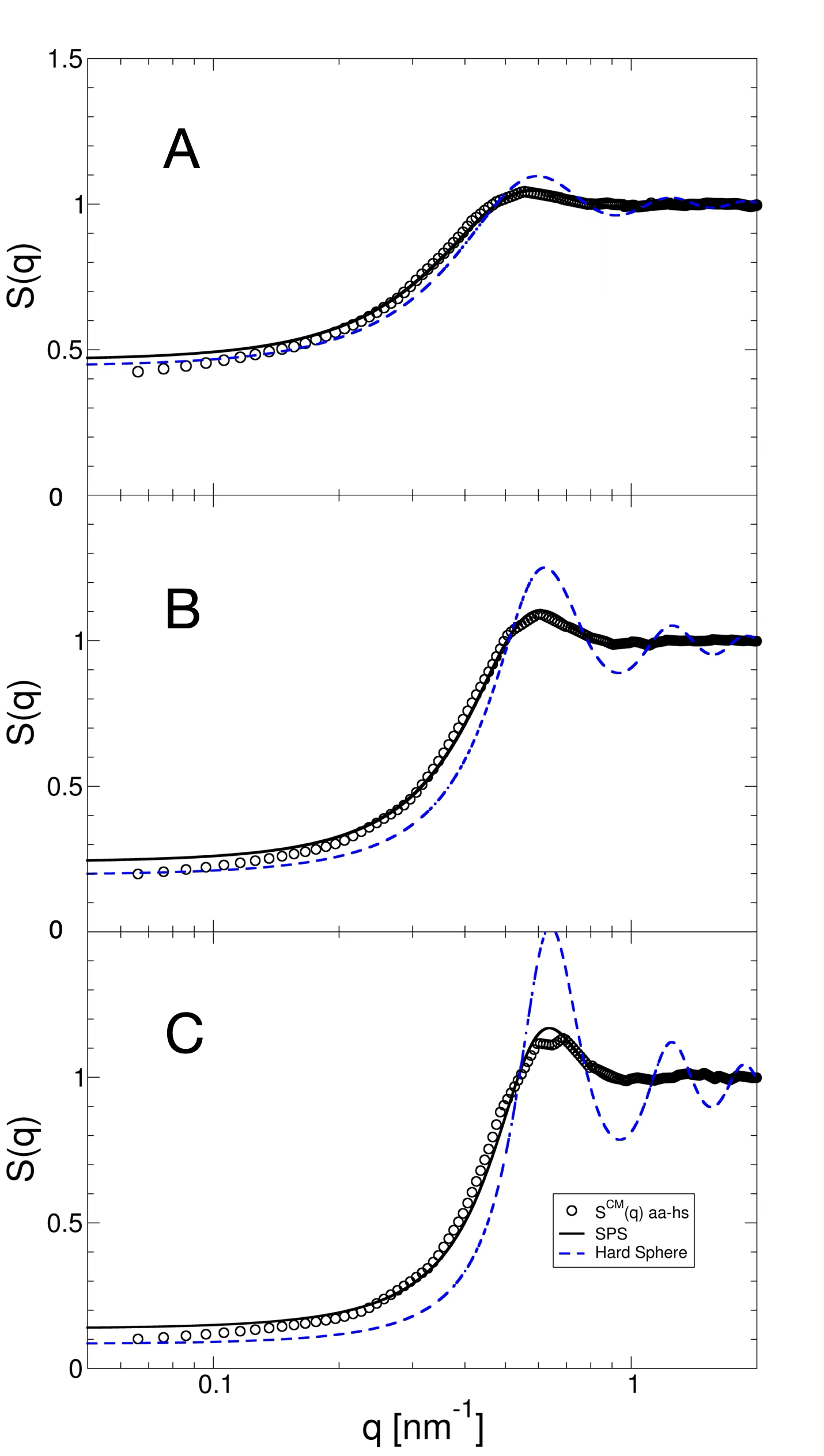}
\caption{Center of mass structure factor $S^{cm}(q)$ for excluded volume interactions. Shown are data from the SPS model (black solid line), the hard sphere colloid model (dashed blue line), and the simulations using the aa-coarse grained model (open black symbols) for three different concentrations. A: 50 mg/ml, 7 mM ionic strength structure; B: 100 mg/ml, 7 mM ionic strength structure; C: 150 mg/ml, 7 mM ionic strength structure.
}
\label{fig:Sq-hsonly}
\end{figure}

Next, we compute the center of mass structure factor, $S^{cm}(q)$, and compare it with the structure factors for the hard sphere and SPS models. 
For monodisperse spherical colloids interacting via an isotropic potential, $S^{cm}(q)$ corresponds to the static structure factor obtained using liquid state theory\cite{Goodstein1975}. Here, the starting point is the link between the static structure factor and the pair distribution function $g(r)$ given by 
 \begin{equation}
	\label{pair-distribution}
	S(q) = 1 + 4 \pi \rho   \int_{0}^{\infty} r^2(g(r) - 1) \frac{\sin qr}{qr} \,dr\
\end{equation}
\noindent where $g(r)$ 
is calculated through an appropriate closure relation, such as HMSA \cite{Zerah1986}. 

Figure \ref{fig:Sq-hsonly} shows the results of these calculations for three representative concentrations in comparison with the results from aa-level coarse-grained simulations. The hard sphere model clearly fails in describing local structural correlations. It strongly overestimates nearest neighbor correlations expressed by the first peak in $S(q)$, occurring at $q^* \approx 2\pi/\sigma$, that is the nearest neighbor distance between two particles. In contrast, the SPS model reproduces the local microstructure almost quantitatively at all concentrations investigated up to relatively large values, above random close packing. Only at low q-values we detect systematic deviations, in agreement with the previously discussed discrepancy between simulations and the SPS model for the osmotic compressibility at high concentrations.

\subsection{Electrostatic and short-range interactions}

\subsubsection{Potential of mean force}

\begin{figure}[!h]
\includegraphics[width=0.9\linewidth]{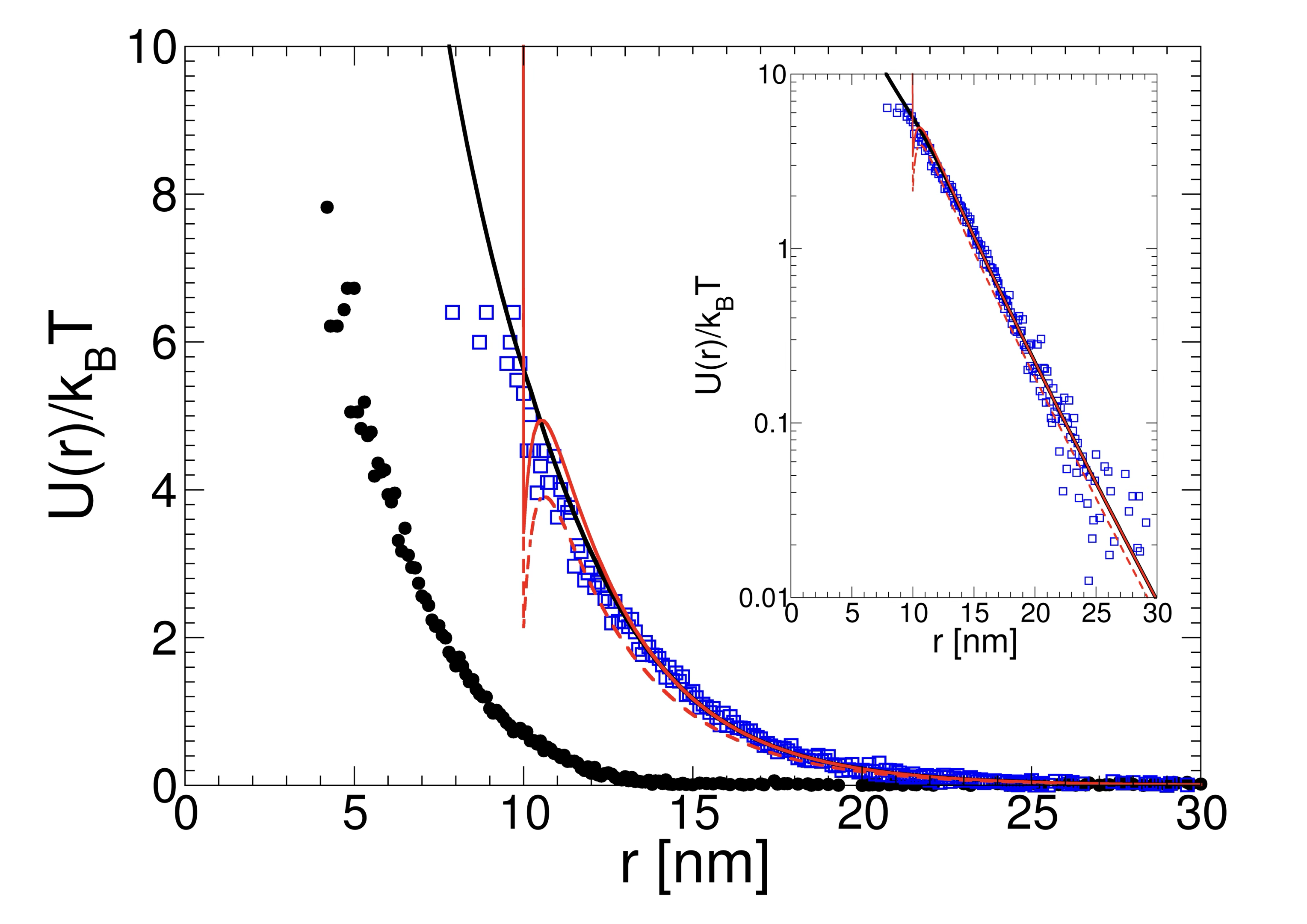}
\caption{Potential of mean force (PMF) as a function of the center-center distance $r$ obtained from MC simulations using an aa coarse-grained model for mAb-1 at 7 mM ionic strength (blue squares). Also shown are the PMF for excluded volume interactions only (black solid symbols), the SPS model with an interaction potential as the solid black line, and that of a colloid model with hard sphere diameter $\sigma_{hs} = 10$ nm and either $Z_{eff} = 21$ as the red dashed line or $Z_{eff} = 23$ as the red solid line, respectively. The inset shows the data from mAb-1, the SPS model, and the colloid model in a log-lin representation.}
\label{fig:PMF7mM}
\end{figure}

Next, we perform simulations to calculate the PMF using the full potential between individual beads in the aa coarse-grained model, as given by eqn. \ref{eq:ham}, which accounts for electrostatic, excluded volume and short-range attractive interactions. The resulting data for mAb-1 at an ionic strength of 7 mM are shown in Fig. \ref{fig:PMF7mM}. A comparison with the PMF for excluded volume effects only (black solid symbols in Fig. \ref{fig:PMF7mM}) shows that interactions are now completely dominated by electrostatics. The potential is much more long-ranged, and for distances $r \gtrsim 13$ nm well described by a screened Coulomb or Yukawa potential. This is also demonstrated in the inset of Fig. \ref{fig:PMF7mM}, where the same data is shown in a log-lin representation. 

It is interesting to compare the PMF obtained from the simulations with the results from a classical colloid model frequently used to interpret results for mAbs~\cite{gulotta2024}. The model is based on an interaction potential $U_{\textrm{coll,t}}(r)$ given by a sum of a repulsive ($U_{\textrm{coll,r}}(r)$) and an attractive ($U_{\textrm{coll,a}}(r)$) term:

\begin{equation}
\begin{aligned}
U_{\textrm{coll,t}}(r) =  U_{\textrm{coll,r}}(r) + U_{\textrm{coll,a}}(r)
\label{eq:Pot-total}
\end{aligned}
\end{equation}

\noindent where the repulsive contribution $U_{\textrm{coll,r}}(r)$ is given by
\begin{equation}
\begin{aligned}
\beta U_{\textrm{coll,r}}(r) = \ & \ \infty, &(r < \sigma_{\textrm{hs}})\\
=\ & \ L_B Z_{eff}^2 \bigg(\frac{e^{\kappa\sigma_{\textrm{hs}}/2}}{1 + \kappa \sigma_{\textrm{hs}}/2}\bigg)^2
 \frac{e^{-\kappa r}}{r},  &(r> \sigma_{\textrm{hs}})
\label{eq:Yukawa}
\end{aligned}
\end{equation}

\noindent with $\sigma_{\textrm{hs}}$ the hard sphere colloid diameter and $Z_{\textrm{eff}}$ the effective charge of the particle, with $\kappa$ given by eqn. \ref{eq:kappa}.

For the attractive term $U_{\textrm{coll,a}}(r)$ we consider an additional short-range attraction, using an approach that has resulted in a quantitative description of the structural properties of concentrated solutions of globular proteins and antibodies at low ionic strength \cite{Cardinaux2007, Cardinaux2011, gulotta2024}. Here $U_{\textrm{coll,a}}(r)$ is given by a power law of the form
\begin{equation}
\begin{aligned}
\beta U_{\textrm{coll,a}}(r) =  - \epsilon_a \bigg( \frac{\sigma_{hs}}{r}\bigg)^\alpha
\label{eq:Pot-att}
\end{aligned}
\end{equation}

\noindent where we use a value $\alpha = 30$. 

The potential given by eqns. \ref{eq:Pot-total} - \ref{eq:Pot-att} has already been used previously to interpret the experimental data for mAb-1, and has resulted in very good agreement with SLS and DLS results using values of $Z_{eff} = 21$ and $\epsilon_a = 3.5$ (see red dashed line in Fig. \ref{fig:PMF7mM}) \cite{gulotta2024}. For an ionic strength of 7 mM and $Z_{eff} = 21$ the attractive term turns out to be irrelevant, as the PMF is completely dominated by the Yukawa term in eqn. \ref{eq:Yukawa}. However, while the long-range part of the potential that determines important physical quantities such as $B_2^*$, $k_D$ and $S(0)$ is indeed well described by the standard charged colloid model, it is important to realize that $Z_{eff} = 21$ is much smaller than the charge  $Z = 31$ of the aa-coarse-grained mAb model used in our computer simulations to generate the PMF shown in Fig. \ref{fig:PMF7mM}. 

The discrepancy between the true and effective charge when using a classical colloid model to describe mAb solutions has already been discussed previously \cite{gulotta2024}. In the colloid model, a mAb is treated as a non-conducting hard sphere with a homogeneous surface charge density, and screening by counterions starts at distances larger than $r > \sigma_{hs}$ as shown in Fig. \ref{CollModel}b. 
When comparing with the actual mAb structure and the location of charges and small ions (Fig.\ref{CollModel}a), a three-arm star polyelectrolyte, such as shown schematically in Fig.\ref{CollModel} (d) bears a much closer resemblance. We therefore test whether we can build a more realistic colloid model based on an SPS model (Fig.\ref{CollModel} (c) ), where the charges and small ions inside the penetrable sphere are distributed analogous to a star polyelectrolyte, and the electrostatic contributions to the PMF are then calculated by the theory for star polyelectrolytes \cite{Denton2003}. 

For the SPS model we start by writing the PMF as a sum of contributions from excluded volume, electrostatic and short range attractive interactions given by

\begin{equation}
\begin{aligned}
U_{\textrm{SPS,t}}(r) =  U_{\textrm{SPS,ev}}(r) + U_{\textrm{SPS,el}}(r) + U_{\textrm{SPS,a}}(r)
\label{eq:Pot-SPS}
\end{aligned}
\end{equation}

\noindent where the excluded volume contribution $U_{\textrm{SPS,ev}}(r)$ is given by eqn. \ref{eq:SPS-ev-pot}.

The term $U_{\textrm{SPS,el}}(r)$ describing the contributions from electrostatic interactions is calculated using an analogy to star polyelectrolytes. A key feature of star polyelectrolytes is that their charge density distribution follows $\rho_c \sim 1/r^2$. Using this assumption, Denton derived an effective PMF that shows two distinct regions, $r < \sigma$ and $r \geq \sigma$, relative to the diameter $\sigma$ of the star polyelectrolyte. We use $\sigma_{SPS}$ as the diameter $\sigma$ of the star polyelectrolyte, i.e. we take the same diameter given by the diameter of the circumscribing sphere of the mAb as used for the calculation of excluded volume effects. 

Following Denton, for $r \geq \sigma_{SPS}$ the electrostatic contribution $U_{\textrm{SPS,el}}(r)$ to $U_{\textrm{SPS,t}}(r)$ is then given by \cite{Denton2003}

\begin{equation}
 \begin{aligned}
\beta U_{\textrm{SPS,el}}(r) 
= Z^2L_B\left(\frac{\text{sinhc}(\kappa a)}{\kappa a}\right)^2 \frac{e^{-\kappa r}}{r}, \ &(r \geq \sigma_{SPS})
\label{eq:Star-Y}
\end{aligned}
\end{equation}

\noindent where $a = \sigma_{SPS}/2$ is the radius of the star-like particle and where
\begin{equation}
 \begin{aligned}
\text{sinhc}(x) \equiv \int_{0}^{x}du \frac{\text{sinh(u)}}{u} = \sum_{n=0}^{\infty}\frac{x^{2n+1}}{(2n+1)(2n+1)!}.
\label{eq:sinhc}
\end{aligned}
\end{equation}

A comparison between the electrostatic interaction potential from the classical colloid model (eqn. \ref{eq:Yukawa}) and that from the 3-arm star polyelectrolyte model (eqn. \ref{eq:Star-Y}) not only shows that both result in a Yukawa potential for distances larger than the diameter of the particle. It also immediately provides us with a relation between the actual net charge $Z$ and the effective charge $Z_{eff}$ that needs to be used in the standard colloid model. We can calculate $Z_{eff}$ by comparing eqns. \ref{eq:Yukawa} and \ref{eq:Star-Y} and assuming that for a given value of the true net charge $Z$ the potential needs to have the same value for both models for distances $r \geq \sigma_{SPS}$. This leads to

\begin{equation}
 \begin{aligned}
L_B Z_{eff}^2 \bigg(\frac{e^{\kappa\sigma_{\textrm{hs}}/2}}{1 + \kappa \sigma_{\textrm{hs}}/2}\bigg)^2= L_BZ^2\left(\frac{\text{sinhc}(\kappa \sigma_{SPS}/2)}{\kappa \sigma_{SPS}/2}\right)^2 ,
\label{eq:Amp}
\end{aligned}
\end{equation}
\noindent which results in the following relationship between net charge $Z$ and effective charge $Z_{eff}$

\begin{equation}
 \begin{aligned}
Z_{eff}= Z \left(\frac{sinhc(\kappa \sigma_{SPS}/2)}{\kappa \sigma_{SPS}/2}\right) /\left(\frac{e^{\kappa\sigma_{\textrm{hs}}/2}}{1 + \kappa \sigma_{\textrm{hs}}/2}\right) .
\label{eq:Zeff}
\end{aligned}
\end{equation}

\noindent Using eqn. \ref{eq:Zeff}, we obtain a value of $Z_{eff} = 23$ for $Z = 31$ and an ionic strength of 7 mM, quite close to the value used previously to reproduce the experimental data obtained for mAb-1 at this ionic strength. A look at Fig. \ref{fig:PMF7mM} shows that the long-range part of the potential is indeed quantitatively described by the colloid model with $Z_{eff} = 23$ (solid red line in Fig. \ref{fig:PMF7mM}). 

In contrast to standard charged colloids, star polyelectrolytes can interpenetrate, and the corresponding effective pair potential $U_{\textrm{\textrm{SPS,el}}}(r)$ for $r < \sigma$ has also been derived by Denton, and is given by a combination of two terms $U_{\textrm{SPS,el}}(r) = u_{mm}(r) + u_{ind}(r)$, characterizing the bare macroion interaction ($u_{mm}(r)$) and the contribution from microion-induced interactions ($u_{ind}(r)$), respectively\cite{Denton2003}. These depend on the degree of overlap and are given by eqns. 26, A3, and A4 in Ref. \cite{Denton2003} . Here we reproduce the set of equations used to describe the effective potential for distances $r < \sigma$,

\begin{equation}
 \begin{aligned}
\beta u_{mm}(r) = \frac{Z^2e^2}{2\epsilon a}[\frac{9}{2}-\frac{7}{4}\frac{r}{a}-\frac{1}{2}\left((3-\frac{a}{r})(1-\frac{r}{a})+\frac{r}{a}\ln (\frac{r}{a}) \right) \\ 
 \times \ln\left(\frac{r-a}{a} \right)+\frac{r}{2a}\int_{1-a/r}^{a/r}dx\frac{\ln x}{1-x}]  ,  a<r\leq2a
\label{eq:umm}
\end{aligned}
\end{equation}

\noindent and

\begin{equation}
 \begin{aligned}
\beta u_{ind}(r) = \frac{16 \pi^2Z^2e^2a^2}{\epsilon r}\int_{0}^{\infty}dx\frac{\sin(xr/a)}{x^3(x^2+\kappa^2a^2)}] 
\label{eq:uind}
\end{aligned}
\end{equation}

\noindent When applying star polyelectrolyte theory for the description of electrostatic interactions in our SPS model, we however also need to realize that the analogy has to break down at short distances, where due to its much bulkier arms the mAb can no longer freely rotate, and where thus the basic assumption of an average charge density distribution given by $\rho_c \sim 1/r^2$ breaks down. We have thus constructed a purely phenomenological relationship that reproduces the PMF obtained in our aa-coarse-grained simulations at different ionic strengths for $\sigma_{core} \leq r \leq 0.7\sigma_{SPS}$:
\begin{equation}
 \begin{aligned}
 \beta U_{SPS,el}(r) = A_{overlap} \left(\exp[2.3(1-r/\sigma_{SPS})(\kappa a)^{0.6}]\right)
\label{eq:pheno}
\end{aligned}
\end{equation}

\noindent where the amplitude $A_{overlap}$ is calculated to match $U_{SPS,el}(r)$ at $r/\sigma_{SPS} = 0.7$ from the star polyelectrolyte expression given by eqns. \ref{eq:umm} and \ref{eq:uind}.

\begin{figure}[!h]
\includegraphics[width=0.9\linewidth]{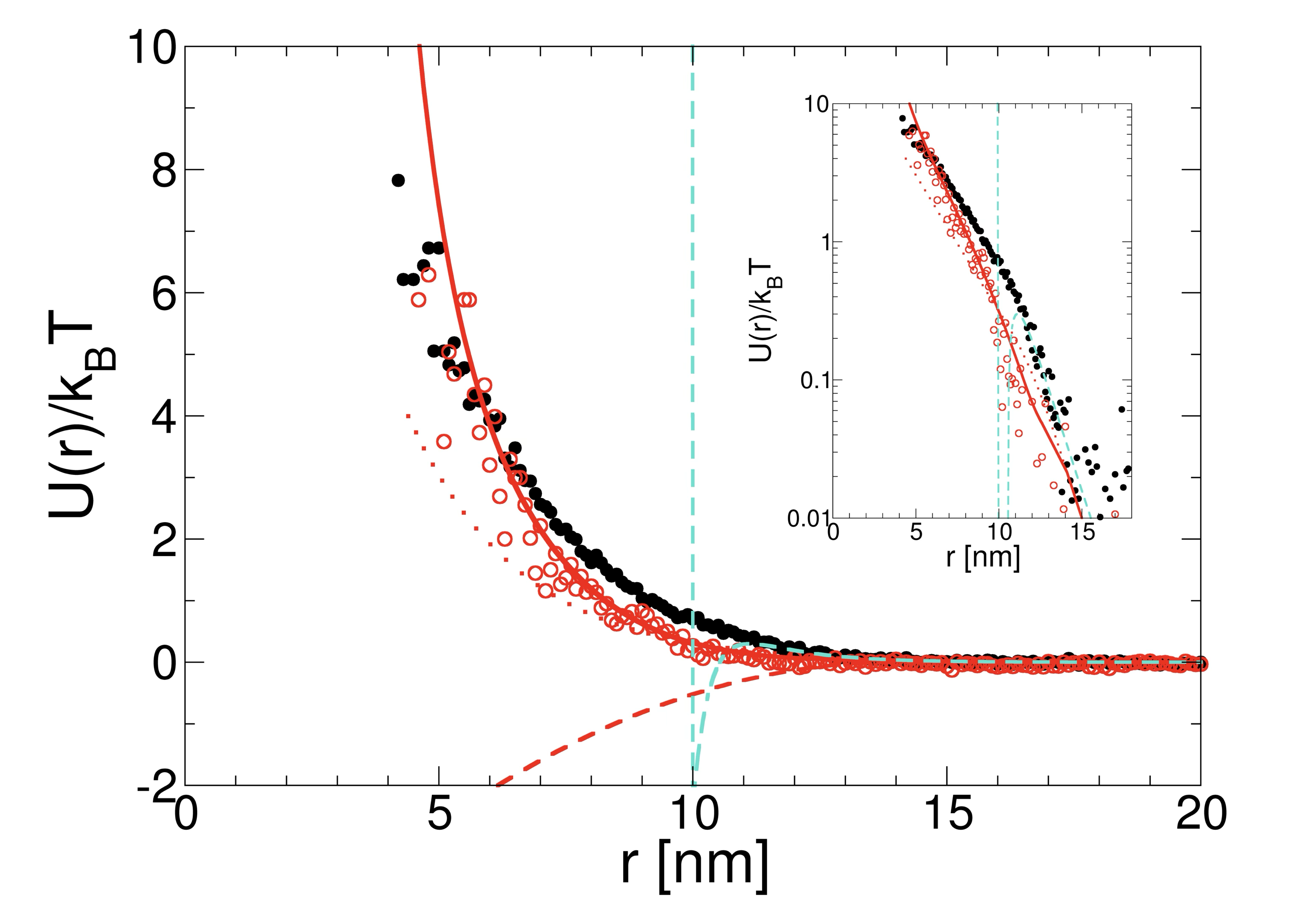}
\caption{Potential of mean force (PMF) as a function of the center-center distance $r$ obtained from MC simulations using an aa coarse grained model for mAb-1 at 57 mM ionic strength with $Z = 34$ (open red circles). Also shown are the calculated PMF for excluded volume interactions only (black solid symbols), the SPS model with $Z = 34$, given by Eq. \ref{eq:Pot-SPS}, as the solid red line, the attractive (Eq.~\ref{eq:star-att}, dashed red line) and electrostatic (Eqs.~\ref{eq:Star-Y}, \ref{eq:umm}-\ref{eq:pheno}, dotted red line) contributions to the SPS model, and that of a colloid model with hard sphere diameter $\sigma_{hs} = 10$ nm and $Z_{eff} = 20$ as the light green dashed line, respectively. The inset shows the same data in a log-lin representation.}
\label{fig:PMF57mM}
\end{figure}

The combination of eqns. \ref{eq:Star-Y} and \ref{eq:umm}-\ref{eq:pheno} agrees very well with the PMF obtained from the aa-coarse-grained simulations at different ionic strengths. This is also demonstrated in Fig. \ref{fig:PMF57mM}, where we plot the PMF for an ionic strength of 57 mM obtained with the aa-coarse-grained simulations and calculated with the SPS model.

At this high ionic strength, Fig. \ref{fig:PMF57mM} shows that the electrostatic contribution $U_{SPS,el}(r)$ is weaker than the excluded volume interactions between mAbs. It is interesting to note that under these conditions the resulting PMF from the aa-coarse-grained simulations is below the pure hard sphere case at intermediate distances $\sigma_{SPS}/2 \lesssim r \lesssim \sigma_{SPS}$, indicating that one also needs to take into account a short range attractive contribution $U_{SPS,a}(r)$ to $U_{SPS,t}(r)$. Such an attractive interaction has been implemented in the aa-coarse-grained simulations using eqn. \ref{eq:ham}, i.e. assuming a short range contribution for each amino acid bead that follows an $r^{-6}$ dependence inspired by van der Waals interactions. While in the classical colloid model $U_{\textrm{coll,a}}(r)$ is often modeled with a short range power law as given by eqn. \ref{eq:Pot-att}, a closer look at the effective PMF from the simulations indicates a much broader, long-range attractive contribution, and we obtain very good agreement between the SPS model and the aa-coarse-grained simulations using 

\begin{equation}
\begin{aligned}
\beta U_{\textrm{SPS,a}}(r) =  - \epsilon_a\left(1 - \frac{r-\sigma_{core}}{\sigma_{SPS} - \sigma_{core}}\right)^{2} 
\label{eq:star-att}
\end{aligned}
\end{equation}

\noindent with $\epsilon_a = 3.0 $.

\subsubsection{Static structure factor}

Having determined the PMF and obtained the corresponding functional forms for a classical colloid model and the SPS model, respectively, we can again calculate $B_2^*$ and the full center-of-mass structure factor $S^{cm}(q)$ for different concentrations and compare them with those obtained directly from the aa-coarse-grained simulations.

\begin{table}[h!]
\centering
\begin{tabular}{| c | c | c | } 
 \hline
   Model & $B_2^*$ [7 mM]  & $B_2^*$ [57 mM] \\ [0.5ex] 
 \hline
   colloid & 5.9 & 0.94   \\ 
  \hline
  SPS & 6.0 & 0.88  \\ [1ex] 
 \hline
\end{tabular}
\caption{The normalised second virial coefficient $B_2^*$ for the classical colloid model (Eqn.~\ref{eq:Pot-total}) and the SPS model (Eqn.~\ref{eq:Pot-SPS}), respectively, where the reference hard sphere system is a hard sphere with diameter $\sigma_{hs} = 10$ nm. Charges used in the calculations are $Z_{eff} = 23.4$ for the colloid model at 7 mM ionic strength and $Z_{eff} = 20.1$ at 57 mM, and $Z = 31$ for the SPS model at 7 mM and $Z = 34$ at 57 mM ionic strength, respectively. The attractive contribution for the corresponding PMF was calculated using $\epsilon = 3.0$ for the SPS model and $\epsilon = 3.5$ for the colloid model.}
\label{tab:B2-comp}
\end{table}

\begin{figure}[h!]
\includegraphics[width=0.95\linewidth]{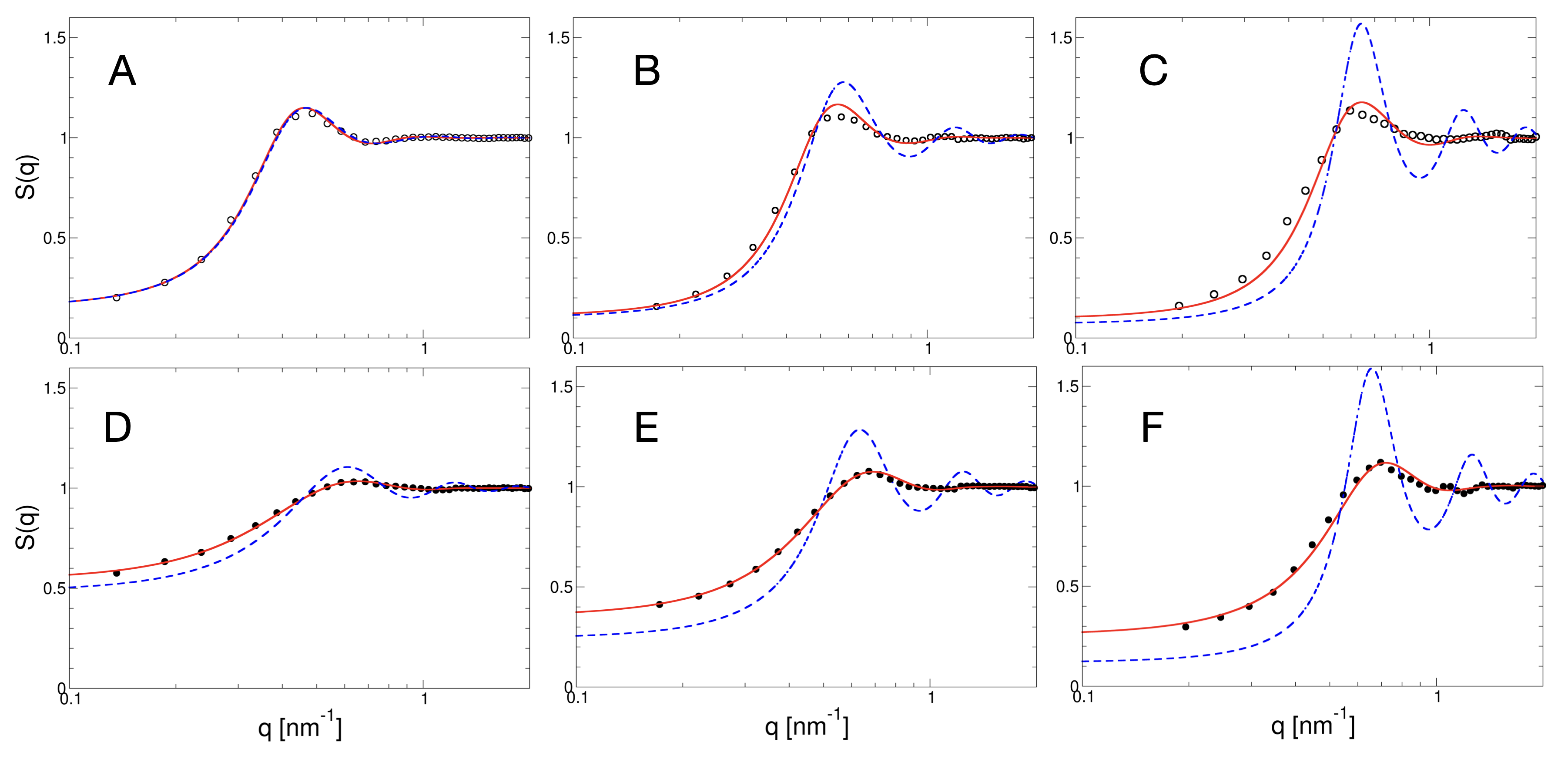}
\caption{Center of mass structure factor $S^{cm}(q)$ as a function of the magnitude of the scattering vector $q$ for several concentrations and two ionic strengths. Shown are data directly obtained from simulations using the aa-coarse-grained model (black symbols) and calculations using liquid state theory (eqn. \ref{pair-distribution}) with the PMF obtained from the SPS model (red solid line) and the classical colloid model (blue dashed line), respectively. Shown are data for an ionic strength of 7 mM (A: $c = 50$ mg/ml, B: $c = 100$ mg/ml, C: $c = 150$ mg/ml) and 57 mM (D: $c = 50$ mg/ml, E: $c = 100$ mg/ml, F: $c = 150$ mg/ml), respectively.}
\label{fig:Sq-sim-hmsa}
\end{figure}

The resulting values for the normalized second virial coefficient $B_2^*$ are summarized in Table \ref{tab:B2-comp}. Both models result in very similar values for $B_2^*$ at the different ionic strengths, indicating that the initial concentration dependence of the osmotic compressibility is the same for both models when using an effective charge $Z_{eff}$ given by eqn. \ref{eq:Zeff} for the colloid model and the `real' charge $Z$ for the SPS model. However, a different picture emerges when we look at the full center-of-mass structure factor $S^{cm}(q)$ at these ionic strength values (Fig. \ref{fig:Sq-sim-hmsa}). While at low ionic strength and low concentration, both models agree quite well and also reproduce the correct $S^{cm}(q)$ obtained from simulations, this is not the case for higher ionic strength and higher concentrations. Here, the classical colloid model predicts much stronger structural correlations than seen in the simulations, with a strongly enhanced nearest neighbor peak and clearly visible higher order peaks. In contrast, the SPS model reproduces the simulation results for $S^{cm}(q)$ almost quantitatively at all conditions investigated.

These differences between the two models are particularly enhanced for high ionic strength, where the major contribution to the PMF arises from excluded volume interactions. Using a simple hard sphere potential as given by eqn.~\ref{eq:harmonic} results in much stronger structural correlations than for the softer contribution seen in simulations and used in the SPS model (eqn. \ref{eq:SPS-ev-pot}). At low ionic strength and low concentrations, the PMF is dominated by the repulsive long-range electrostatic interactions, and the differences between both models become almost negligible when using an appropriately rescaled effective charge $Z_{eff}$ as given by eqn. \ref{eq:Zeff}. It is quite remarkable that the use of a centrosymmetric potential of mean force based on soft penetrable particles, as implemented in the SPS model, is capable of reproducing the structural correlations observed in simulations for the highly asymmetric aa-coarse-grained model of a monoclonal antibody.

\subsection{Comparison with experimental data}

\subsubsection{mAb-1: Osmotic compressibility and apparent hydrodynamic radius}

We first test the SPS model with experimental data for mAb-1, and focus on the osmotic compressibility or $S(0)$ and the apparent hydrodynamic radius. The experimental data have previously been published and were then interpreted using a classical colloid model\cite{gulotta2024}. Here, we compare the experimental data with the predictions from the SPS model and a colloid model using the parameters given in the caption of Table \ref{tab:B2-comp}. In the colloid model, we apply the standard colloid potential given by eqns. \ref{eq:Pot-total} - \ref{eq:Pot-att}, and calculate $Z_{eff}$ from the true charge $Z$ based on the comparison with star polyelectrolyte theory (eqn. \ref{eq:Zeff}). In the calculations of the corresponding PMFs, we have to take into account that the Debye length $\lambda_D$ depends on the mAb concentration \cite{gulotta2024}. Here, we use an ad-hoc expression for the screening parameter $\kappa(c)$ in order to include the changes in ion concentrations induced by the ultrafiltration step performed for making the high concentration stock solution as a result of the Gibbs-Donnan equilibrium, followed by the dilution with the buffer \cite{gulotta2024}.
This results in an implicit dependence of the potential of mean force on the mAb concentration, as shown in Fig.S5 in SI for the SPS model. Moreover, in the hybrid colloid model, the effective charge is also concentration dependent, as shown in Fig. S6 in the SI.

\begin{figure}[h!]
\includegraphics[width=0.9\linewidth]{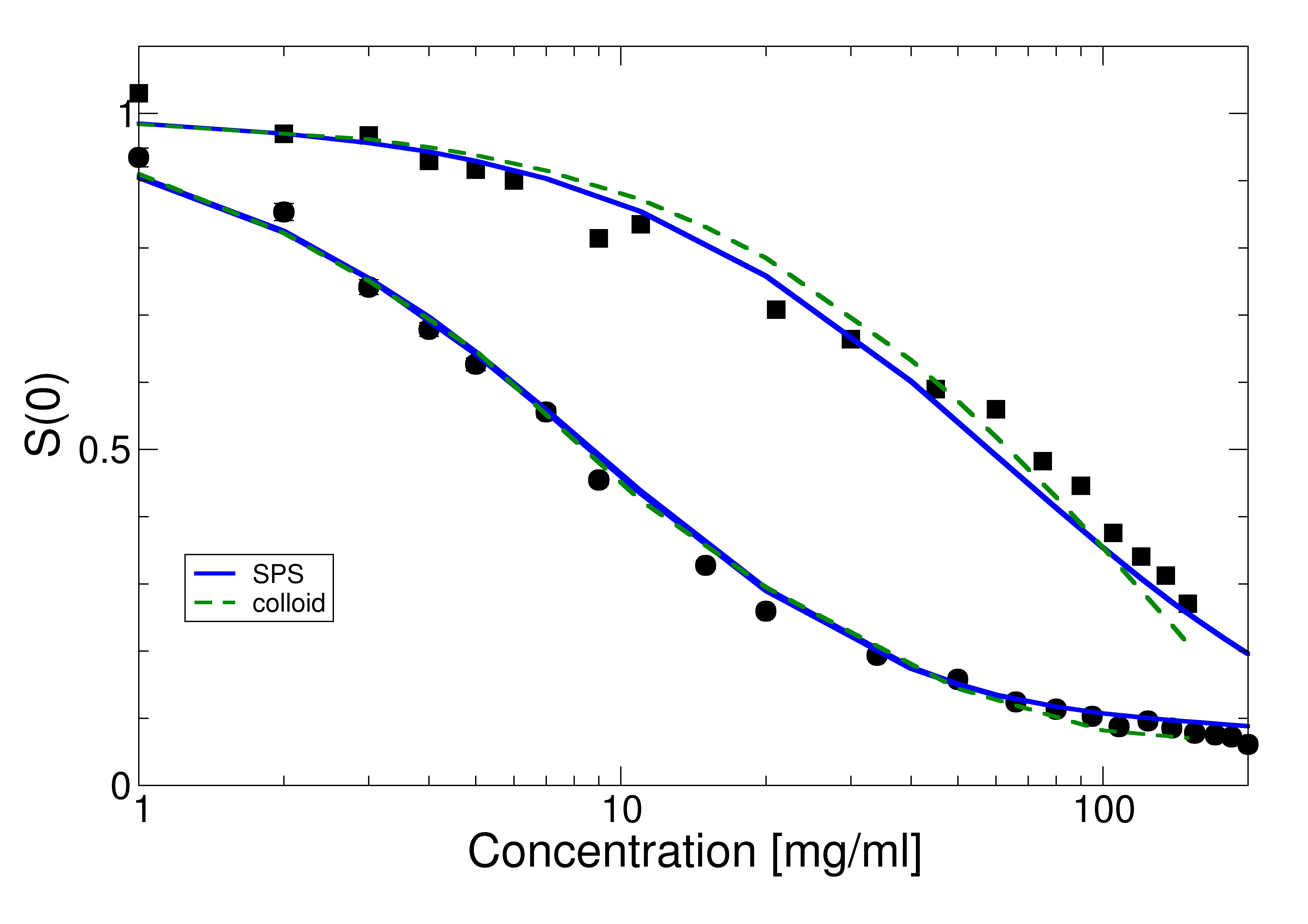}
\caption{A comparison of the measured $S(0)$ vs. \emph{c} obtained for mAb-1 at two ionic strengths (7 mM and 57 mM) with different models. Shown are the experimental data (7 mM: solid black circles; 57 mM: solid black squares, data taken from ref. \cite{gulotta2024}), and the theoretical results for the SPS model using the HMSA closure relation (blue solid line) and the colloid model with $Z_{eff}$ determined for each concentration using eqn. \ref{eq:Zeff} (dashed green line). }
\label{fig:S0-mAb-1}
\end{figure}

Using the corresponding expressions for the PMF, we then calculate $S(0)$ with the HMSA closure relation. The results for the different models are shown in Fig. \ref{fig:S0-mAb-1}. For 7 mM ionic strength, the SPS and the colloid model are virtually identical up to concentrations of $c \approx 50$ mg/ml. For the higher ionic strength of 57 mM, the differences are slightly larger, but overall comparable or smaller than the noise in the experimental data. At a concentration of around 50 - 70 mg/ml there appears to be a characteristic change in the relative slope of the classical colloid model compared to the SPS model, which is reminiscent of the split in the concentration dependence of $S(0)$ for pure excluded volume interactions for both models as shown in Fig. \ref{fig:S0-hs}. It is, however, clear from Fig. \ref{fig:S0-mAb-1} that the osmotic compressibility can be adequately reproduced with both models, provided that an effective charge $Z_{eff}$ as given by eqn. \ref{eq:Zeff} is used for the colloid model. 

\begin{figure}[h!]
\includegraphics[width=0.9\linewidth]{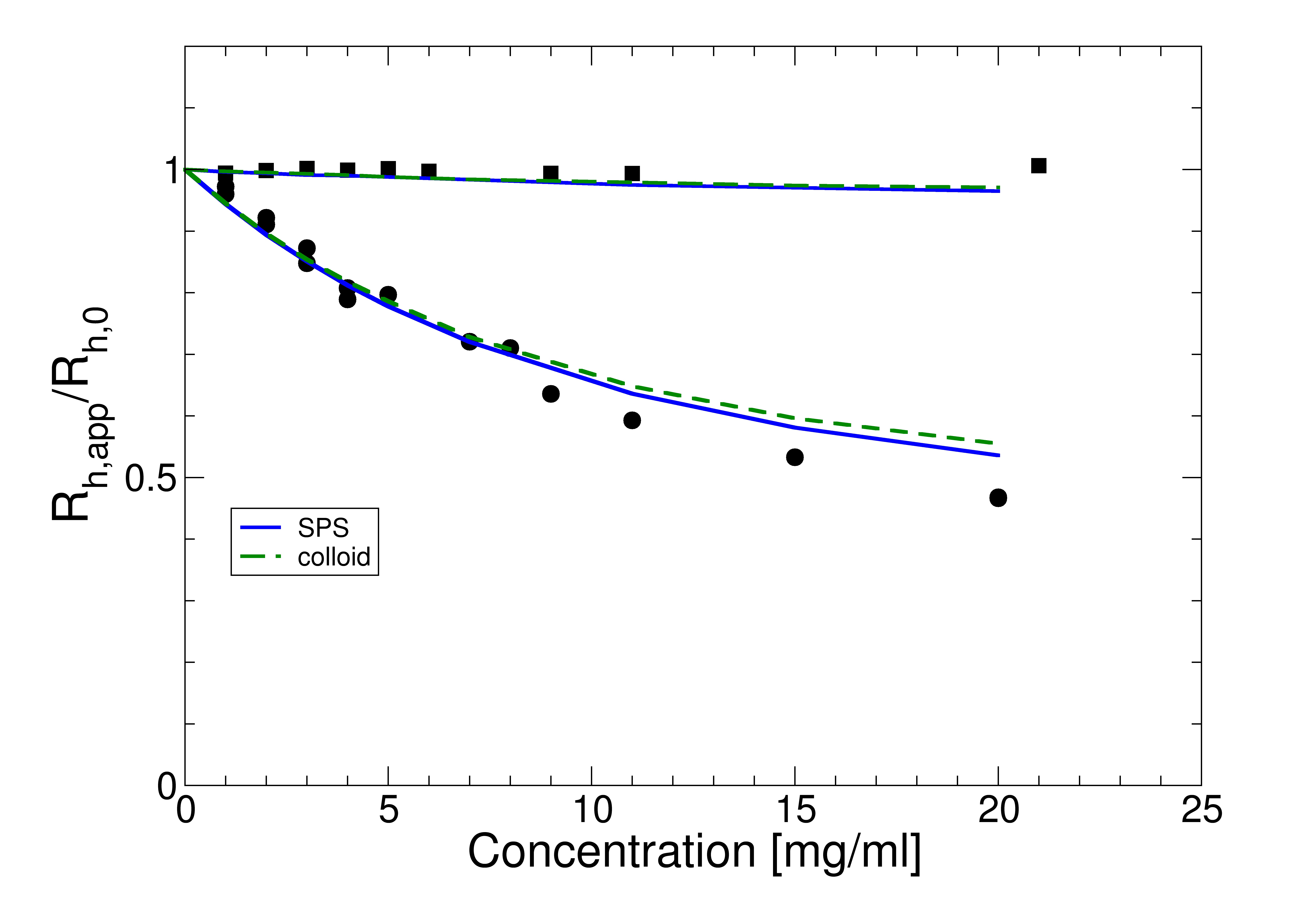}
\caption{A comparison of the measured normalized hydrodynamic radius $R_\mathrm{h,app}/R_{\mathrm{h},0}$ vs. \emph{c} obtained for mAb-1 at two ionic strengths (7 mM and 57 mM) with different models. Shown are the experimental data (7 mM: solid black circles; 57 mM: solid black squares, data taken from ref. \cite{gulotta2024}), and the theoretical results for the SPS model (blue solid line), and the colloid model with $Z_{eff}$ determined for each concentration using eqn. \ref{eq:Zeff} 
(dashed green line). }
\label{fig:Rhapp-mAb-1}
\end{figure}

We can also compare the experimental data for the apparent hydrodynamic radius $R_\mathrm{h,app}$ shown in Fig. \ref{fig:Rhapp-mAb-1}  with the predictions from the SPS and colloid models, respectively. $R_\mathrm{h,app}$ is calculated using the relationship between the short-time collective diffusion coefficient $D_\mathrm{c}^\mathrm{s}(q)$ and the ideal diffusion coefficient $D_0$ in the absence of interactions, given by \cite{Naegele1996, Banchio2008}
 \begin{align}
	\label{Dcoll}
	D_\mathrm{c}^\mathrm{s}(q) = D_0 \frac{H(q)}{S(q)}
\end{align}
where $H(q)$ is the hydrodynamic function that describes the effects of hydrodynamic interactions, and where $S(q)$ is calculated using HMSA as described above. For the calculation of $H(q)$ we assume pairwise additive hydrodynamic interactions, which should be accurate up to $c \lesssim 25$ mg/ml. We can thus write \cite{Neal1999}
\begin{align}
	\label{Hq}
	H(q) = 1 + 6 \pi \rho R_\mathrm{h}  \int_{0}^{\infty} r(g(r) - 1) F(qr) \,dr\ .
\end{align}
where $F(qr)$ is given by
\begin{align}
	\label{Fqr}
    F(qr) = \bigg[ \frac{\sin qr}{qr} +  \frac{\cos qr}{(qr)^2} -  \frac{\sin qr}{(qr)^3}\bigg].
    \end{align}

For small particles such as proteins, the measured diffusion coefficient corresponds to the so-called gradient diffusion coefficient given by 
\begin{align}
	\label{Dc}
	D_\mathrm{c} = \lim_{q \to 0} D_\mathrm{c}^\mathrm{s}(q) = D_0 \frac{H(0)}{S(0)} 
\end{align}
\noindent where $H(0)$ is related to the sedimentation velocity, $U_\mathrm{sed}$ \cite{Banchio2008}. 

This results in

\begin{align}
	\label{Rhapp}
	\frac{D_0}{D_\mathrm{c}} = \frac{R_\mathrm{h,app}}{R_{\mathrm{h},0}} =  \frac{S(0)}{H(0)} 
\end{align}
\noindent where $R_{\mathrm{h},0}$ is the hydrodynamic radius of the mAb in the absence of interactions, i.e., at infinite dilution. In the calculations, we use the $S(0)$ obtained with the HMSA closure relation for the corresponding model, and the $g(r)$ also from HMSA for the calculation of $H(0)$. The resulting data for both models are also shown in Fig. \ref{fig:Rhapp-mAb-1}. The values for the colloid model are almost indistinguishable from those obtained with the SPS model when using a $Z_{eff}(c)$ calculated with eqn. \ref{eq:Zeff}, 
 and both models agree quite well with the experimental data.

\subsubsection{mAb-1: Static structure factor}

\begin{figure}[h!]
\includegraphics[width=0.9\linewidth]{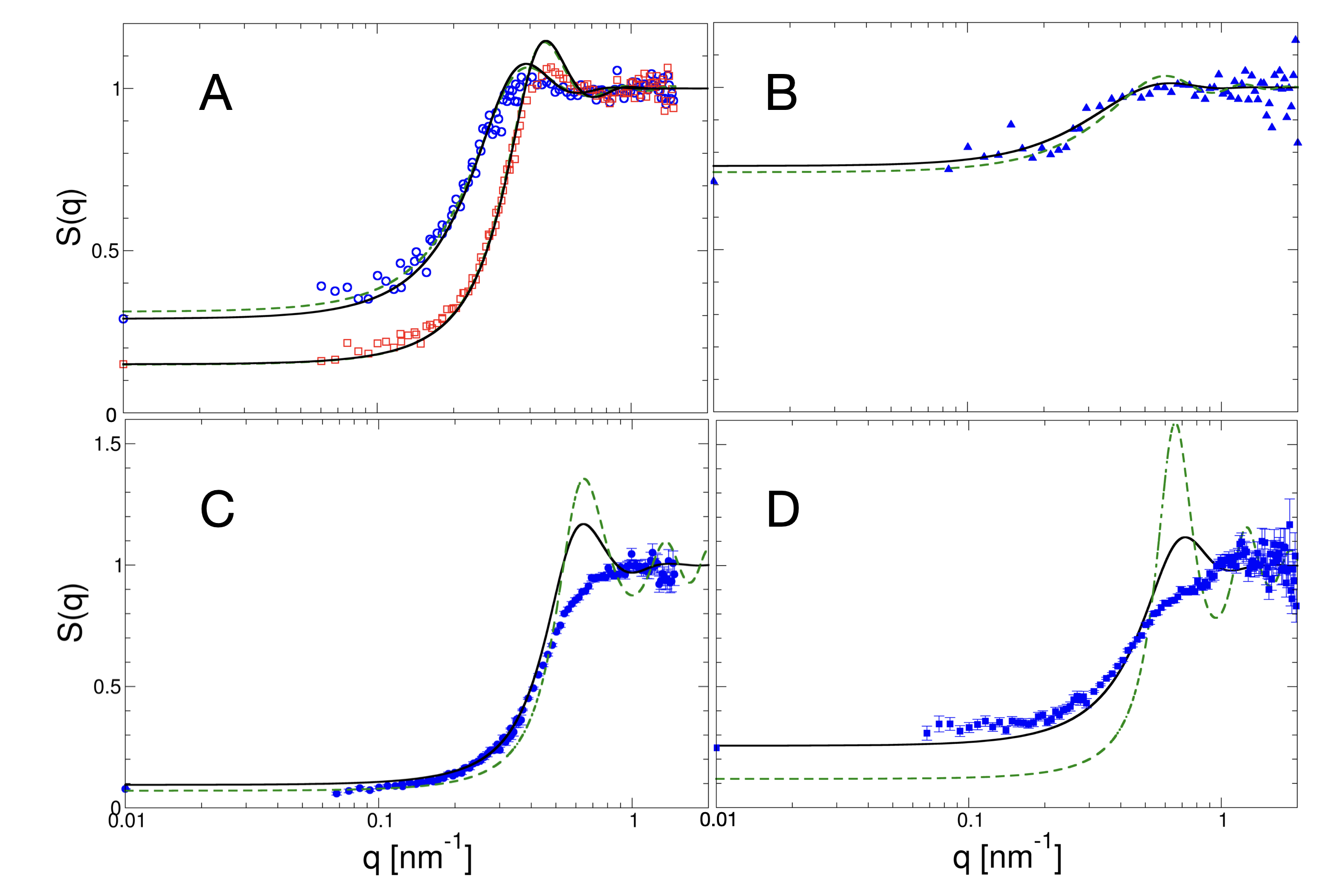}
\caption{Measured effective structure factor $S(q)$ vs. $q$ obtained for mAb-1 at two ionic strengths (7 mM and 57 mM) compared with calculated structure factors $S(q)$ for the colloid and SPS models. Shown are the experimental data (A: 7 mM ionic strength, 20 mg/ml (open blue circles), 50 mg/ml (open red squares); B: 57 mM ionic strength, 20 mg/ml (solid blue triangles); C: 7 mM ionic strength, 150 mg/ml (solid blue circles); D: 57 mM ionic strength, 150 mg/ml (solid blue circles), data taken from ref. \cite{Polimeni2024}), and the theoretical results for the SPS model (black solid line), and the colloid model with $Z_{eff}$ determined for each concentration using eqn. \ref{eq:Zeff} 
(dashed green line). Also shown are the experimental values for $S(0)$ obtained with SLS for each sample at $q = 0.01$ nm$^{-1}$ taken from ref. \cite{gulotta2024} }
\label{fig:Sq-eff}
\end{figure}

While the experimental data for the compressibility and the apparent hydrodynamic radius are well reproduced by the SPS and the colloid model when using a $Z_{eff}(c)$ calculated with eqn. \ref{eq:Zeff}
, the situation changes when looking at the full measured effective structure factor obtained from SAXS (Fig. \ref{fig:Sq-eff}). At low ionic strength and low concentrations $c \lesssim 50$ mg/ml, both models are in good agreement with the measured data, although the nearest neighbor peak is slightly overestimated at higher concentrations. Moreover, both models result in almost identical structure factors, in particular at low ionic strength, where the long-range contributions from electrostatic interactions dominate.

At higher concentrations both models fail, and the nearest neighbor peak is strongly overestimated. This failure to reproduce the measured or effective structure factor is particularly evident for the colloid model and at higher ionic strength, as shown in Fig. \ref{fig:Sq-eff}D. The fact that spherical particle models are not able to reproduce measured structure factors for anisotropic particles such as mAbs is well known. The relationship $S^{cm}(q) = S_{eff}(q)$ is only correct for monodisperse spherical particles, whereas this is not the case for anisotropic objects \cite{gulotta2024}. Here, the total scattering intensity measured in a scattering experiment can no longer be described by independent contributions from particle shape (particle form factor $P(q)$) and interparticle correlation effects (structure factor $S^{cm}(q)$). In fact, for anisotropic objects, the scattering intensity depends on the orientation of the particle, and for interacting particles the orientation between particle pairs at distances closer or smaller than their overall diameter is no longer uncorrelated or random. There have been attempts to overcome this problem and use approximate schemes such as the decoupling approximation (DCA) given by 
\begin{equation}
\label{decoupling}
  S^{DCA}_{eff}(q) = 1 + \beta(q)[S^{cm}(q) - 1]
\end{equation}
\noindent where $\beta(q) = \left\langle \lvert F(\bf q) \rvert \right\rangle^2 / \left\langle \lvert F(\bf q) \rvert ^2 \right\rangle$ and $F(\bf q)$ is the orientation-dependent scattering amplitude of an anisotropic object \cite{Yearley2013, Corbett2017, Pedersen2001}.

\begin{figure}[h!]
\includegraphics[width=0.9\linewidth]{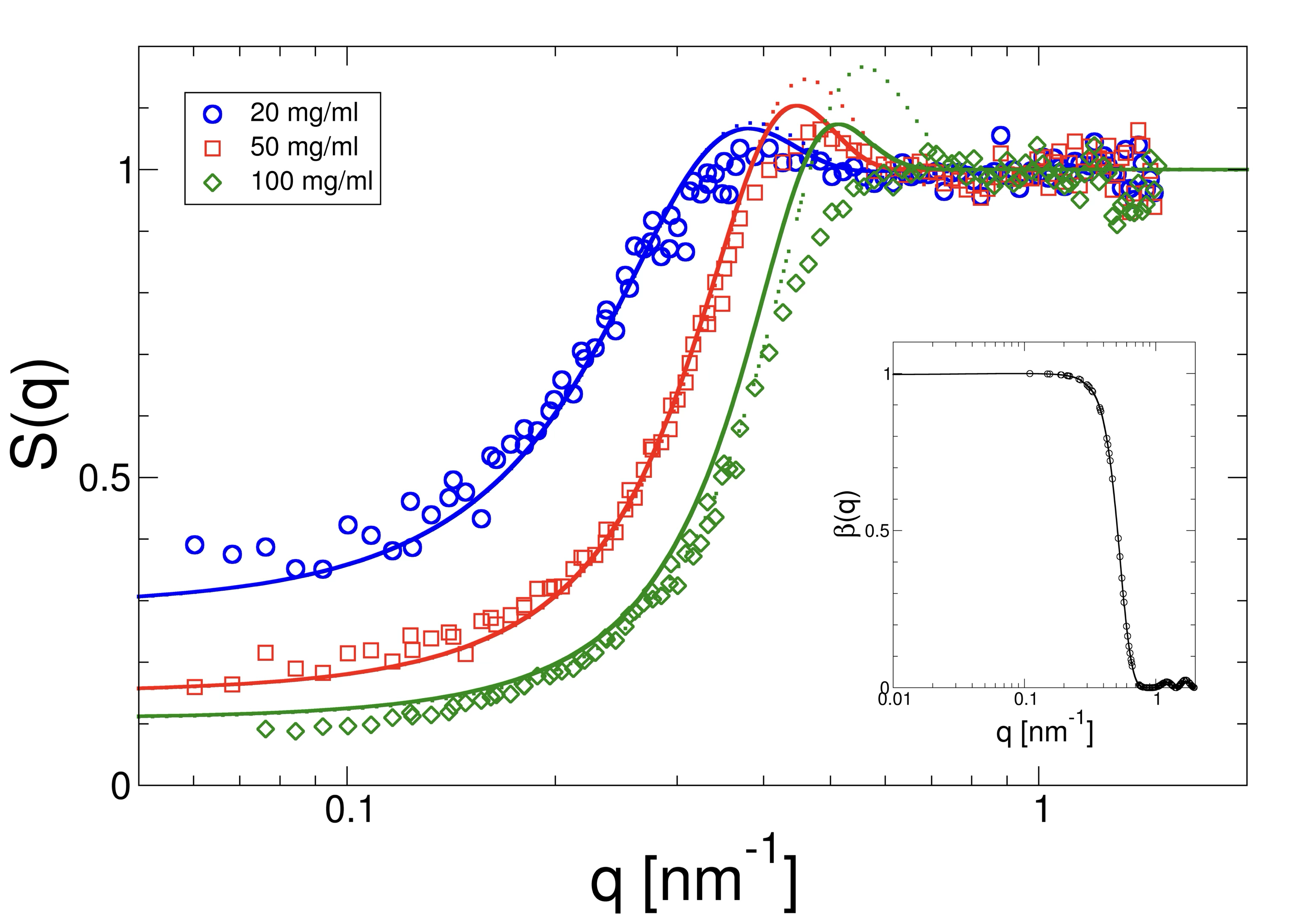}
\caption{Decoupling approximation applied to the center-of-mass structure factor $S^{cm}(q)$ obtained for mAb-1 at 7 mM ionic strength, calculated using HMSA for the SPS model. Shown are the experimentally measured $S(q)$ for 20 mg/ml (open blue circles), 50 mg/ml (open red squares) and 100 mg/ml (open green diamonds) (Data taken from ref. \cite{Polimeni2024}), and the theoretical results for $S^{cm}(q)$ (dotted lines) and the results from the decoupling approximation $S^{DCA}_{eff}(q)$ calculated using eqn. \ref{decoupling}. Also shown is $\beta(q)$, calculated from the aa-coarse-grained model of mAb-1 as an inset.}
\label{fig:Sq-DCA}
\end{figure}

As shown in Fig. \ref{fig:Sq-DCA}, the DCA indeed improves the agreement at low ionic strength and concentrations below $c \leq 50$ mg/ml. At higher concentrations, however, the DCA also fails and clearly overestimates the nearest neighbor peak that is almost absent experimentally. A key assumption in the DCA is that particle orientations and positions are not correlated. While this is justified at low ionic strength and lower concentrations, where the long-range electrostatic repulsion keeps particles away at distances larger than their overall diameter, this is no longer the case at higher concentrations and/or high ionic strength. For concentrations $c \gtrsim 100$ mg/ml, i.e., above the overlap concentration $ c^* \approx 95$ mg/ml, we expect that individual mAbs can no longer freely rotate, thus violating the underlying assumption of the decoupling approximation. This is further aggravated by the fact that the ionic strength also increases with increasing concentration, thus making the potential of mean force less long-ranged. This enhances the probability for particle distances shorter than the overall mAb diameter, thus also making the assumption of uncorrelated particle orientation and position less justified. In principle, the SPS model would be well suited for an application of the decoupling approximation due to the fact that the model reproduces the correct center-of-mass structure factor at all concentrations and ionic strengths investigated. However, the large shape anisotropy reduces the applicability of the DCA to relatively low concentrations $c \lesssim 50$ mg/ml, where the differences between $S^{DCA}_{eff}(q)$ and $S^{cm}(q)$ are not very large. Nevertheless, in this concentration range, the SPS model is indeed able to reproduce measured structure factors obtained by SAXS, and thus allows for a determination of the correct net charge $Z$ of mAbs with a relatively homogeneous charge distribution.

\subsubsection{Linear virial regime, mAb-1 and mAb-2}

When characterising mAb solutions, the interaction coefficients $k_I$ and $k_D$ are often used to probe protein-protein interactions. These interaction coefficients describe the concentration dependence of $S(0)$ and $R_{h,app}$ at low concentrations, i.e., in the linear virial regime, which is usually expressed by
\begin{align}
	\label{eqn:kI}
	S(0) = 1 - k_\mathrm{I} c
\end{align}
and
\begin{align}
	\label{eqn:kD}
	R_\mathrm{h,app} = R_{\mathrm{h},0} (1 - k_\mathrm{D} c),
\end{align}
where the concentration $c$ is given in mg/ml \cite{Corti1981}. $k_I$ and $k_D$ provide us with an overall description of the potential of mean force $U(r)$. $k_I$ is related to the second virial coefficient $B_2$ by
\begin{align}
	\label{eqn:kI-B2}
	k_\mathrm{I} = 2 \frac{N_\mathrm{A}}{M_\mathrm{w}}B_2
\end{align}
with $N_\mathrm{A}$ being Avogadro's number and $B_2$ given by eqn. \ref{B2}. We can thus directly obtain $k_I$ from the PMF through

 \begin{equation}
	\label{eq:kI-calc}
	k_I = \frac{4\pi N_A}{M} \pi  \int_{0}^{\infty}  \bigg(1 - e^{-\beta U(r)}\bigg) r^2 \,dr\
\end{equation}

In order to derive an expression for $k_D$, we start with eqn. \ref{Rhapp}. In the limit of low concentrations, i.e., in the linear virial regime, we approximate the pair correlation function $g(r)$ with $g(r) \approx$ exp$[-\beta U(r)]$ and rewrite eqn. \ref{Hq}
\begin{align}
	\label{Hqkf}
	H(0) = \lim_{q \to 0}\left\{1 + 6 \pi \rho R_\mathrm{h} \int_{0}^{\infty}  r \left[e^{-\beta U(r)} - 1\right] F(qr) \,dr\right\} 
\end{align}
We then rewrite Eq. \eqref{Hqkf} as
\begin{align}
	\label{H0approx}
	H(0) \approx 1 - k_\mathrm{f} c 
\end{align}
where the interaction coefficient $k_\mathrm{f}$ is given by
\begin{align}
	\label{eq:kf}
k_\mathrm{f} = 6 \pi R_\mathrm{h} \frac{N_\mathrm{A}}{M_\mathrm{w}} \int_{0}^{\infty}  r \left[1 - e^{-\beta U(r)}\right] F(qr) \,dr .
\end{align}
Using Eq. \eqref{eqn:kD} we can then calculate $k_\mathrm{D}$ from
\begin{align}
	\label{eq:Rhapprox}
	\frac{R_\mathrm{h,app}}{R_{\mathrm{h},0}} = \frac{S(0)}{H(0)} \approx \frac{1-k_\mathrm{I}c}{1-k_\mathrm{f}c} \approx 1-(k_\mathrm{I}-k_\mathrm{f})c \approx 1 - k_\mathrm{D}c
\end{align}
where $k_\mathrm{D} = k_\mathrm{I} - k_\mathrm{f}$.

We can use eqns. \ref{eq:kI-calc}, \ref{eq:kf} and \ref{eq:Rhapprox} in order to calculate $k_I$ and $k_D$ for different mAbs and compare the predictions of the SPS and the colloid model with experimental results obtained at different solvent conditions. The results for the model calculations are summarized in Table \ref{tab:kI-kD-comp} for mAb-1. Given the typical errors encountered when measuring $k_I$ and $k_D$ with static and dynamic light scattering, both models are virtually identical, provided that one uses star polyelectrolyte theory to calculate the effective charge $Z_{eff}$ for a given value of the net charge $Z$ and the ionic strength (eqn. \ref{eq:Zeff}). 

\begin{table}[h!]
\centering
\begin{tabular}{| c | c | c | c |  c | } 
 \hline
        & SPS & SPS & colloid & colloid \\
   Model & $k_I$ [cm$^3$/g]  & $k_D$ [cm$^3$/g] & $k_I$ [cm$^3$/g] & $k_D$ [cm$^3$/g] \\ [0.5ex] 
 \hline
   mAb-1, 7 mM & 102.4 & 60.2 & 102.2  &  60.1   \\ 
  \hline
     mAb-1, 57 mM & 15.0 & 3.0 & 15.7 &  2.7    \\ 
 \hline
\end{tabular}
\caption{The interaction coefficients $k_I$ and $k_D$ for the SPS and the colloid model calculated for mAb-1 and ionic strengths 7 mM and 57 mM, respectively. Charges used in the calculations are $Z_{eff} = 23.4$ for the colloid model at 7 mM ionic strength and $Z_{eff} = 20.1$ at 57 mM, and $Z = 31$ for the SPS model at 7 mM and $Z = 34$ at 57 mM ionic strength, respectively. The attractive contribution for the corresponding PMF was calculated using $\epsilon = 3.0$ for the SPS model and $\epsilon = 3.5$ for the colloid model.}
\label{tab:kI-kD-comp}
\end{table}

\begin{figure}[h!]
\centering
\begin{subfigure}{0.49\textwidth}
\centering
\includegraphics[width = \textwidth]{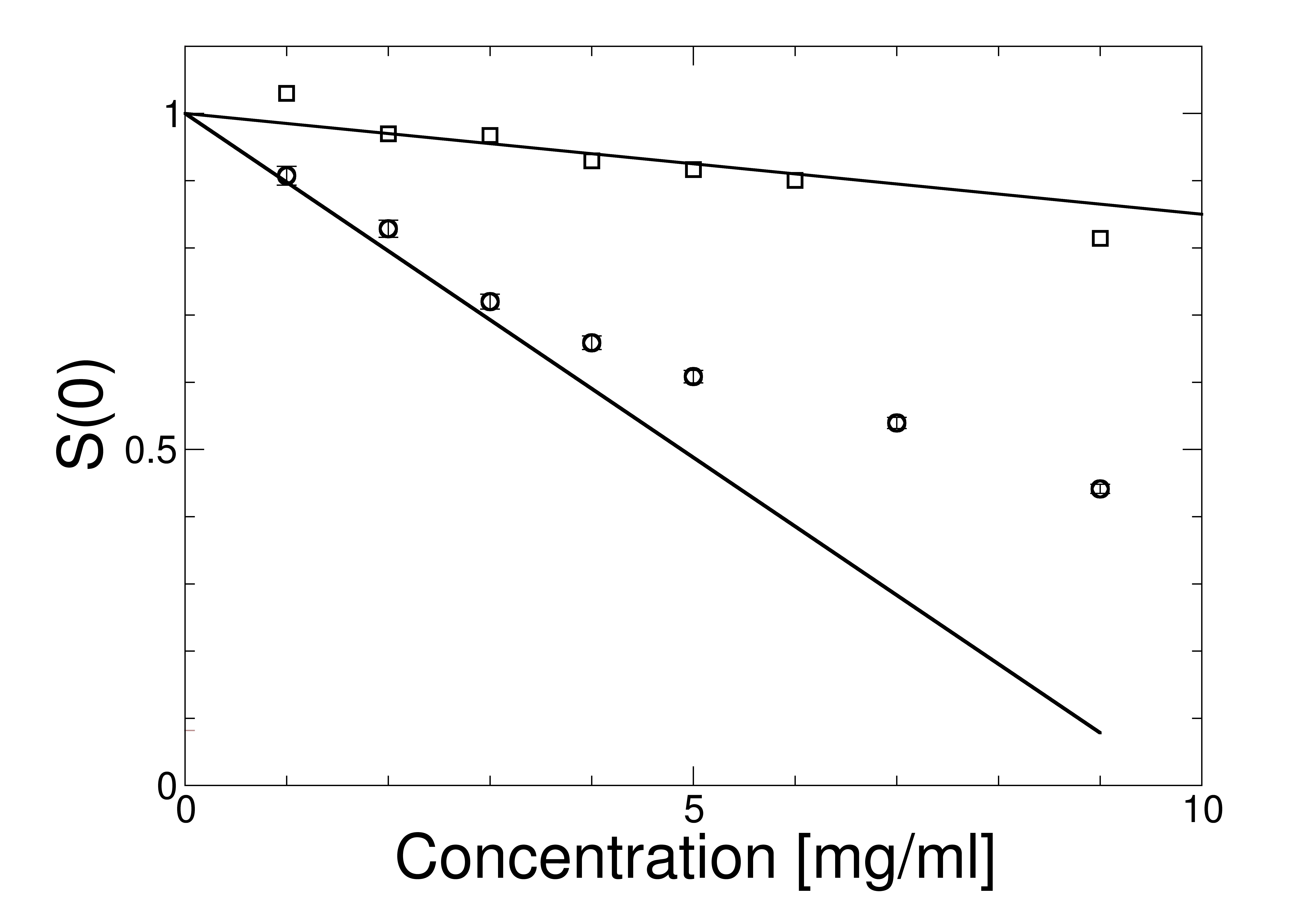}
\caption{}
\end{subfigure}
\begin{subfigure}{0.49\textwidth}
\centering
\includegraphics[width = \textwidth]{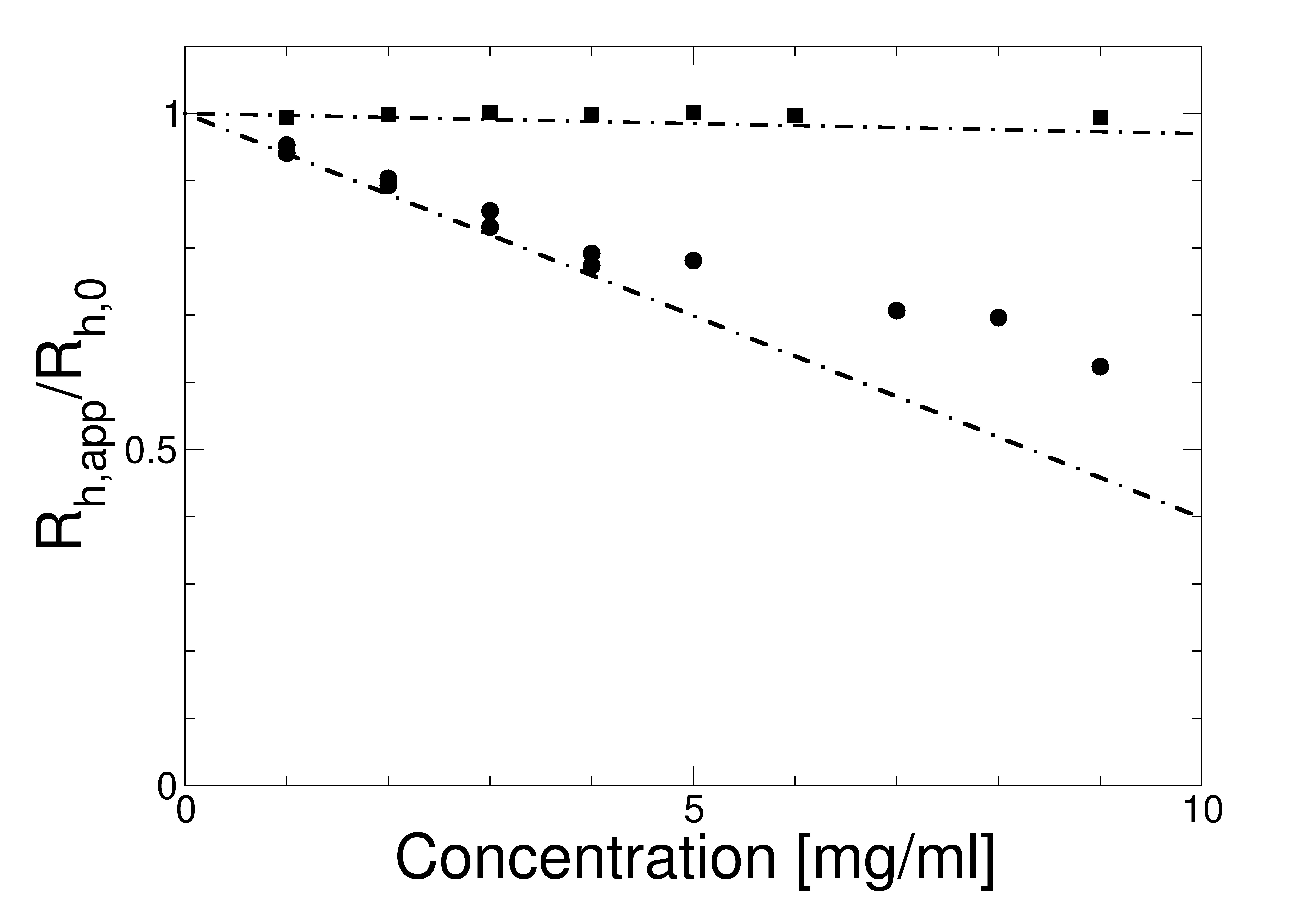}
\caption{}
\end{subfigure}
\caption{Initial concentration dependence of $S(0)$ (left) and $\frac{R_\mathrm{h,app}}{R_{\mathrm{h},0}}$ (right) for mAb-1 in 7 mM (black circles) and 57 mM (black squares) ionic strength compared with predictions for the linear virial regime calculated for the SPS model (black lines). Experimental data taken from ref. \cite{Polimeni2024} .}
\label{fig:mAbG-kd-kI}
\end{figure}

For mAb-1, the experimentally measured initial concentration dependence of $S(0)$ and $\frac{R_\mathrm{h,app}}{R_{\mathrm{h},0}}$ is well reproduced by the SPS model for both ionic strengths investigated, but Fig. \ref{fig:mAbG-kd-kI} also demonstrates that the linear regime can be very narrow for low ionic strength and high protein charges, thus making experimental determinations of $k_I$ and $k_d$ difficult and prone to systematic errors.

\begin{figure}[h!]
\centering
\begin{subfigure}{0.65\textwidth}
\centering
\includegraphics[width = \textwidth]{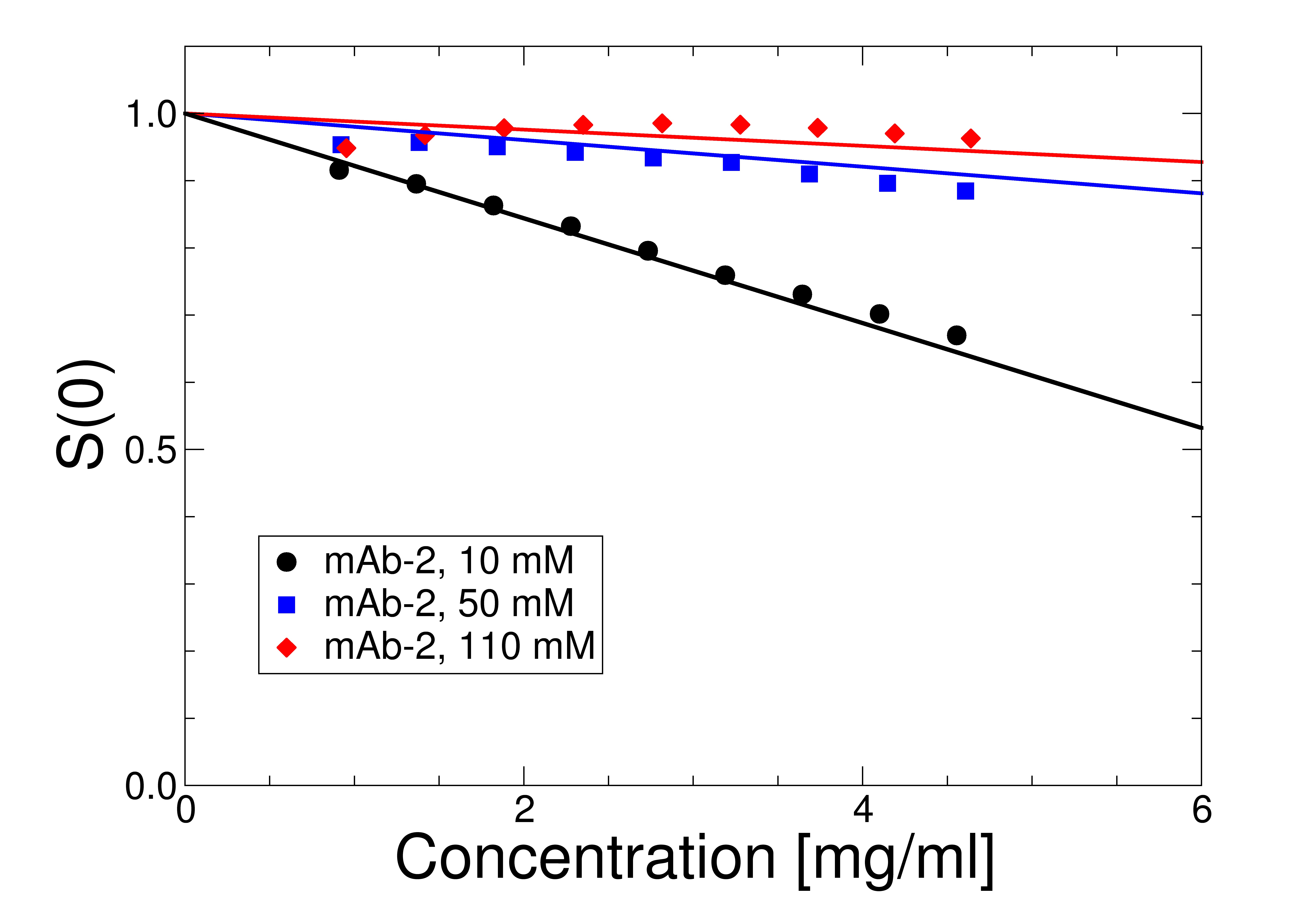}
\end{subfigure}
\begin{subfigure}{0.65\textwidth}
\centering
\includegraphics[width = \textwidth]{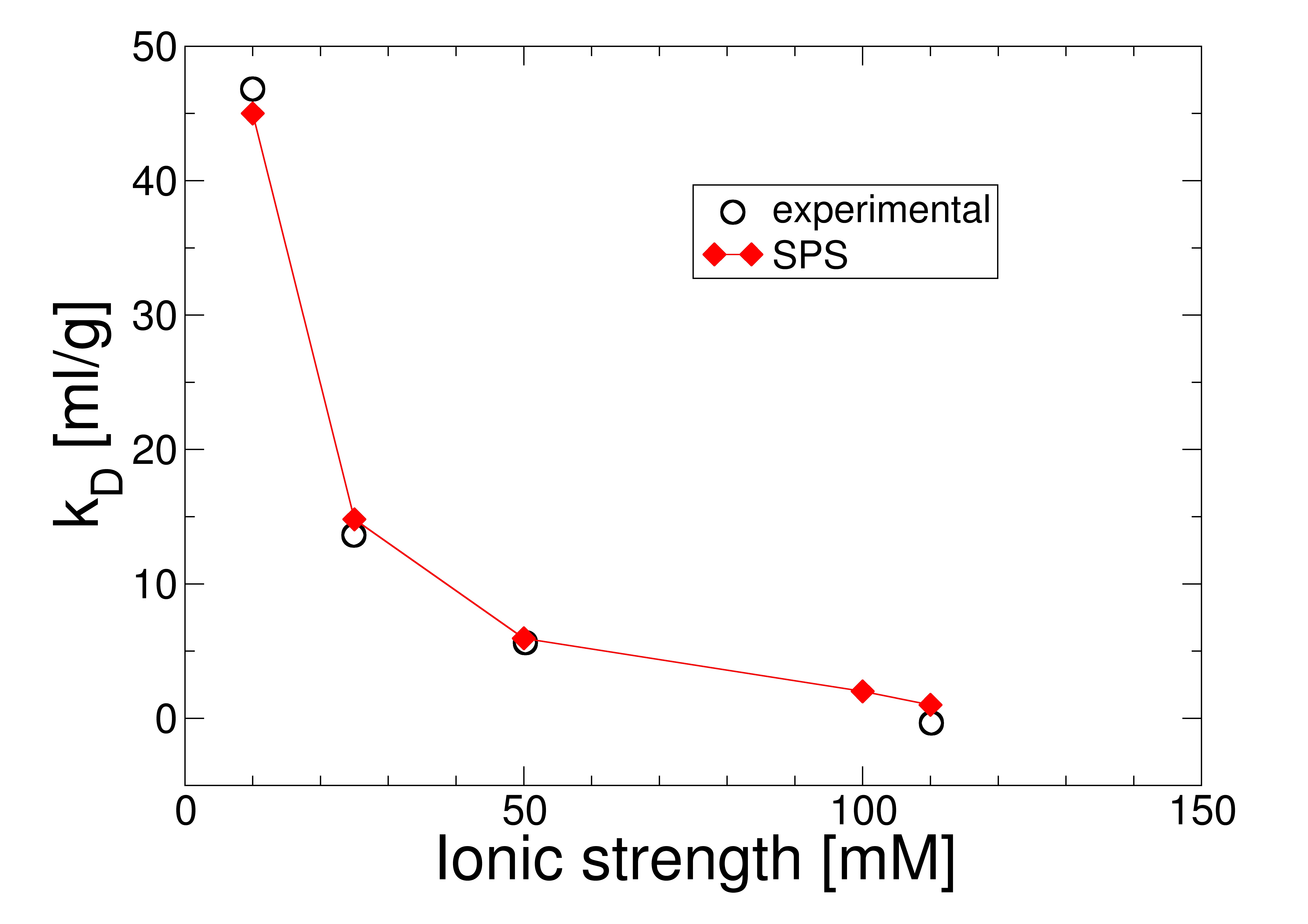}
\end{subfigure}
\caption{Initial concentration dependence of $S(0)$ (top) and ionic strength dependence of $k_D$ (bottom) for mAb-2. Shown are experimental data for $S(0)$ at 10 mM (black circles), 50 mM (blue squares), and 110 mM (red diamonds) ionic strength compared with predictions for the linear virial regime calculated for the SPS model (solid lines). The bottom graph shows the experimental data for $k_D$ as a function of the ionic strength (open black circles) and the predictions for the SPS model (red diamonds).}
\label{fig:NISTmAb-kd-kI}
\end{figure}

We have also compared the SPS model predictions for $k_I$ and $k_D$ with experimental data obtained for mAb-2. As shown in Fig. S2 in SI, at pH = 5.0, this mAb also has a rather homogeneous charge distribution with an average positive net charge of $Z = 35$ at 10 mM, $Z = 39$ at 50 mM, and $Z = 40$ at 110 mM ionic strength as determined from constant-pH simulations. When using these values for the charge, adjusting the diameter $\sigma_{SPS} = 15.5$ nm to the slightly larger overall dimension of the mAb as judged from the size of the circumscribing sphere, and leaving the attractive well depth unchanged ($\epsilon = 3.0$), the SPS model reproduces the experimental data very well as demonstrated in Fig. \ref{fig:NISTmAb-kd-kI}. It is worth pointing out that this agreement has been achieved without any free parameters, as all input parameters required by the SPS model have been directly taken from the initial aa-coarse-grained model simulations ($Z(I,pH)$ as a function of ionic strength and pH; overall diameter $\sigma_{SPS}$) or left unchanged from the initial construction of the model with data from mAb-1 ($\epsilon_a$, $U_{SPS,ev}$).

\section{Conclusions}

Colloid models have frequently been used to describe the influence of protein-protein interactions on antibody solution properties. However, it is already known that when classical colloid models are used, some key parameters - such as the net charge assumed by the model - are effective values that cannot be determined directly from the known molecular structure of a specific mAb.  \cite{gulotta2024}. Thus, these models lack predictive power and can only be used to interpret and rationalize experimental observations. Moreover, the use of a hard sphere interaction to describe excluded volume interactions between mAbs strongly overestimates local correlation effects at high concentrations as demonstrated in this study (Figs. \ref{fig:Sq-hsonly} and \ref{fig:Sq-sim-hmsa}).

The Soft Penetrable Sphere (SPS) model described here overcomes these shortcomings. The use of a soft excluded volume potential as given by eqn. \ref{eq:SPS-ev-pot} quantitatively reproduces the center of mass structure factors $S^{cm}(q)$ obtained with computer simulations based on our weakly coarse-grained amino acid level bead model at all concentrations investigated. Another important improvement comes from the treatment of electrostatic interactions based on the analogy to the well-established theory for star polyelectrolytes, motivated by the similarities between the overall shape of a 3-arm star polyelectrolyte and a mAb. This has allowed us to construct a potential of mean force that quantitatively describes the PMF from computer simulations at low and high ionic strength. 

The SPS model has allowed us to better understand the discrepancy between the actual net charge on the mAb and the effective charge $Z_{eff}$ that needs to be used in the classical colloid model to correctly describe electrostatic interactions between mAbs. The star polyelectrolyte analogy has, in fact, allowed us to calculate $Z_{eff}$ for different ionic strengths and for a given mAb net charge that can, for example, be determined through constant-pH simulations. If one is only interested in low concentration solution properties ($k_I$, $k_D$) or overall thermodynamic or collective dynamic properties such as $S(0)$ or $\frac{R_\mathrm{h,app}}{R_{\mathrm{h},0}}$, one can use a classical colloid model and calculate the corresponding values for $Z_{eff}$ using eqn. \ref{eq:Zeff} for each concentration and ionic strength. As demonstrated in Figs. \ref{fig:S0-mAb-1} and \ref{fig:Rhapp-mAb-1}, the resulting errors are within experimental accuracy when compared to the full SPS model. However, we expect that the use of the SPS model has significant advantages when looking at more microscopic quantities on length scales of the nearest neighbor distance, where details of the local structure, i.e., the correct description of the pair correlation function, matter. This includes, for example, the full dynamic structure factor $S(q,t)$ measured with quasielastic neutron scattering or X-ray photon correlation spectroscopy.

In this work, a soft penetrable sphere model is shown to improve the representation of antibody–antibody interactions. We further speculate that this framework may be particularly well suited to describing conditions where interactions are weakly attractive in the absence of significant electrostatic repulsion. Previous studies have shown that short-range isotropic attractive models can capture thermodynamic properties up to moderate concentrations, but only when using an effective particle diameter smaller than the excluded volume\cite{Lanzaro2021,Hung2019a}. This non-physical adjustment suggests that such models implicitly account for particle softness or interpenetration, which is more naturally captured within a penetrable sphere description.

However, there remain several deficiencies that are inherent to models based on an isotropic potential of mean force. First of all, while we can correctly calculate the center of mass static structure factor $S^{cm}(q)$ or the pair correlation function $g(r)$ for the strongly anisotropic mAbs using the SPS PMF and the HMSA closure relation, we cannot reproduce the effective structure factor measured with SAXS. At lower concentrations, the use of a decoupling approximation improves the results, but this fails at higher concentrations where individual mAbs start to overlap. Furthermore, while the SPS model has predictive power for mAbs with a rather homogeneous charge distribution, it is expected to fail for highly heterogeneous charge distributions where oppositely charged patches result in additional electrostatic attractions and thus an additional orientation-dependent contribution to the overall potential. It will be interesting to see whether we can treat the influence of such patchy contributions by incorporating an additional term from an orientation-averaged pair potential between dipolar particles.\cite{Bratko2002}

\begin{acknowledgement}
We gratefully acknowledge financial support by the Swedish Research Council (VR; Grant Nos. 2019-06075 and 2022-03142). 
The computer simulations were enabled by resources provided by the National Academic Infrastructure for Supercomputing in Sweden (NAISS) and the Swedish National Infrastructure for Computing (SNIC) at Lund University, partially funded by the Swedish Research Council through grant agreements no. 2022-06725 and no. 2018-05973. 
The NISTmAb was provided courtesy of the National Institutes of Standards and Technology in Gaithersburg, MD, USA.
This work is part of the “LINXS Antibodies in Solution” research program and we acknowledge the financial support by the LINXS Institute of Advanced Neutron and X-ray Science.
\end{acknowledgement}

\begin{suppinfo}

Additional information about the charge distribution of mAb-1 and mAb-2, a schematic description of the simulations for obtaining the potential of mean force and for the many-mAb simulations, the concentration dependence of the potential of mean force in the SPS model and of the effective charge in the colloid model.

\end{suppinfo}

\bibliography{mAb-bib}

\providecommand{\latin}[1]{#1}
\makeatletter
\providecommand{\doi}
  {\begingroup\let\do\@makeother\dospecials
  \catcode`\{=1 \catcode`\}=2 \doi@aux}
\providecommand{\doi@aux}[1]{\endgroup\texttt{#1}}
\makeatother
\providecommand*\mcitethebibliography{\thebibliography}
\csname @ifundefined\endcsname{endmcitethebibliography}
  {\let\endmcitethebibliography\endthebibliography}{}
\begin{mcitethebibliography}{44}
\providecommand*\natexlab[1]{#1}
\providecommand*\mciteSetBstSublistMode[1]{}
\providecommand*\mciteSetBstMaxWidthForm[2]{}
\providecommand*\mciteBstWouldAddEndPuncttrue
  {\def\EndOfBibitem{\unskip.}}
\providecommand*\mciteBstWouldAddEndPunctfalse
  {\let\EndOfBibitem\relax}
\providecommand*\mciteSetBstMidEndSepPunct[3]{}
\providecommand*\mciteSetBstSublistLabelBeginEnd[3]{}
\providecommand*\EndOfBibitem{}
\mciteSetBstSublistMode{f}
\mciteSetBstMaxWidthForm{subitem}{(\alph{mcitesubitemcount})}
\mciteSetBstSublistLabelBeginEnd
  {\mcitemaxwidthsubitemform\space}
  {\relax}
  {\relax}

\bibitem[Muschol and Rosenberger(1997)Muschol, and
  Rosenberger]{muschol1997liquid}
Muschol,~M.; Rosenberger,~F. Liquid--liquid phase separation in supersaturated
  lysozyme solutions and associated precipitate formation/crystallization.
  \emph{The Journal of chemical physics} \textbf{1997}, \emph{107},
  1953--1962\relax
\mciteBstWouldAddEndPuncttrue
\mciteSetBstMidEndSepPunct{\mcitedefaultmidpunct}
{\mcitedefaultendpunct}{\mcitedefaultseppunct}\relax
\EndOfBibitem
\bibitem[Woldeyes \latin{et~al.}(2017)Woldeyes, Calero-Rubio, Furst, and
  Roberts]{woldeyes2017predicting}
Woldeyes,~M.~A.; Calero-Rubio,~C.; Furst,~E.~M.; Roberts,~C.~J. Predicting
  protein interactions of concentrated globular protein solutions using
  colloidal models. \emph{The Journal of Physical Chemistry B} \textbf{2017},
  \emph{121}, 4756--4767\relax
\mciteBstWouldAddEndPuncttrue
\mciteSetBstMidEndSepPunct{\mcitedefaultmidpunct}
{\mcitedefaultendpunct}{\mcitedefaultseppunct}\relax
\EndOfBibitem
\bibitem[Stradner and Schurtenberger(2020)Stradner, and
  Schurtenberger]{stradner2020potential}
Stradner,~A.; Schurtenberger,~P. Potential and limits of a colloid approach to
  protein solutions. \emph{Soft Matter} \textbf{2020}, \emph{16},
  307--323\relax
\mciteBstWouldAddEndPuncttrue
\mciteSetBstMidEndSepPunct{\mcitedefaultmidpunct}
{\mcitedefaultendpunct}{\mcitedefaultseppunct}\relax
\EndOfBibitem
\bibitem[Neal \latin{et~al.}(1999)Neal, Asthagiri, Velev, Lenhoff, and
  Kaler]{Neal1999}
Neal,~B.; Asthagiri,~D.; Velev,~O.; Lenhoff,~A.; Kaler,~E. {Why is the osmotic
  second virial coefficient related to protein crystallization?} \emph{Journal
  of Crystal Growth} \textbf{1999}, \emph{196}, 377--387\relax
\mciteBstWouldAddEndPuncttrue
\mciteSetBstMidEndSepPunct{\mcitedefaultmidpunct}
{\mcitedefaultendpunct}{\mcitedefaultseppunct}\relax
\EndOfBibitem
\bibitem[Madani \latin{et~al.}(2025)Madani, Hamacher, and Platten]{Madani2025}
Madani,~M.; Hamacher,~T.; Platten,~F. {Urea and salt induced modulation of
  protein interactions: implications for crystallization and liquid-liquid
  phase separation}. \emph{Soft Matter} \textbf{2025}, \emph{21},
  1937--1948\relax
\mciteBstWouldAddEndPuncttrue
\mciteSetBstMidEndSepPunct{\mcitedefaultmidpunct}
{\mcitedefaultendpunct}{\mcitedefaultseppunct}\relax
\EndOfBibitem
\bibitem[Foffi \latin{et~al.}(2014)Foffi, Savin, Bucciarelli, Dorsaz, Thurston,
  Stradner, and Schurtenberger]{Foffi2014}
Foffi,~G.; Savin,~G.; Bucciarelli,~S.; Dorsaz,~N.; Thurston,~G.~M.;
  Stradner,~A.; Schurtenberger,~P. {Hard sphere-like glass transition in eye
  lens $\alpha$-crystallin solutions}. \emph{Proceedings of the National
  Academy of Sciences} \textbf{2014}, \emph{111}, 16748--16753\relax
\mciteBstWouldAddEndPuncttrue
\mciteSetBstMidEndSepPunct{\mcitedefaultmidpunct}
{\mcitedefaultendpunct}{\mcitedefaultseppunct}\relax
\EndOfBibitem
\bibitem[Bucciarelli \latin{et~al.}(2015)Bucciarelli, Casal-Dujat, {De
  Michele}, Sciortino, Dhont, Bergenholtz, Farago, Schurtenberger, and
  Stradner]{Bucciarelli2015}
Bucciarelli,~S.; Casal-Dujat,~L.; {De Michele},~C.; Sciortino,~F.; Dhont,~J.;
  Bergenholtz,~J.; Farago,~B.; Schurtenberger,~P.; Stradner,~A. {Unusual
  Dynamics of Concentration Fluctuations in Solutions of Weakly Attractive
  Globular Proteins}. \emph{Journal of Physical Chemistry Letters}
  \textbf{2015}, \emph{6}, 4470--4474\relax
\mciteBstWouldAddEndPuncttrue
\mciteSetBstMidEndSepPunct{\mcitedefaultmidpunct}
{\mcitedefaultendpunct}{\mcitedefaultseppunct}\relax
\EndOfBibitem
\bibitem[Bucciarelli \latin{et~al.}(2016)Bucciarelli, Myung, Farago, Das,
  Vliegenthart, Holderer, Winkler, Schurtenberger, Gompper, and
  Stradner]{Bucciarelli2016}
Bucciarelli,~S.; Myung,~J.~S.; Farago,~B.; Das,~S.; Vliegenthart,~G.~A.;
  Holderer,~O.; Winkler,~R.~G.; Schurtenberger,~P.; Gompper,~G.; Stradner,~A.
  {Dramatic influence of patchy attractions on short-time protein diffusion
  under crowded conditions}. \emph{Science Advances} \textbf{2016}, \emph{2},
  e1601432\relax
\mciteBstWouldAddEndPuncttrue
\mciteSetBstMidEndSepPunct{\mcitedefaultmidpunct}
{\mcitedefaultendpunct}{\mcitedefaultseppunct}\relax
\EndOfBibitem
\bibitem[Bergman \latin{et~al.}(2025)Bergman, Garting, {De Michele},
  Schurtenberger, and Stradner]{Bergman2025}
Bergman,~M.~J.; Garting,~T.; {De Michele},~C.; Schurtenberger,~P.; Stradner,~A.
  {Dynamical arrest for globular proteins with patchy attractions}. \emph{Soft
  Matter} \textbf{2025}, \emph{21}, 1152--1161\relax
\mciteBstWouldAddEndPuncttrue
\mciteSetBstMidEndSepPunct{\mcitedefaultmidpunct}
{\mcitedefaultendpunct}{\mcitedefaultseppunct}\relax
\EndOfBibitem
\bibitem[Fusco and Charbonneau(2016)Fusco, and Charbonneau]{fusco2016soft}
Fusco,~D.; Charbonneau,~P. Soft matter perspective on protein crystal assembly.
  \emph{Colloids and Surfaces B: Biointerfaces} \textbf{2016}, \emph{137},
  22--31\relax
\mciteBstWouldAddEndPuncttrue
\mciteSetBstMidEndSepPunct{\mcitedefaultmidpunct}
{\mcitedefaultendpunct}{\mcitedefaultseppunct}\relax
\EndOfBibitem
\bibitem[McManus \latin{et~al.}(2016)McManus, Charbonneau, Zaccarelli, and
  Asherie]{mcmanus2016physics}
McManus,~J.~J.; Charbonneau,~P.; Zaccarelli,~E.; Asherie,~N. The physics of
  protein self-assembly. \emph{Current opinion in colloid \& interface science}
  \textbf{2016}, \emph{22}, 73--79\relax
\mciteBstWouldAddEndPuncttrue
\mciteSetBstMidEndSepPunct{\mcitedefaultmidpunct}
{\mcitedefaultendpunct}{\mcitedefaultseppunct}\relax
\EndOfBibitem
\bibitem[G{\"o}gelein \latin{et~al.}(2008)G{\"o}gelein, N{\"a}gele, Tuinier,
  Gibaud, Stradner, and Schurtenberger]{gogelein2008simple}
G{\"o}gelein,~C.; N{\"a}gele,~G.; Tuinier,~R.; Gibaud,~T.; Stradner,~A.;
  Schurtenberger,~P. A simple patchy colloid model for the phase behavior of
  lysozyme dispersions. \emph{The Journal of chemical physics} \textbf{2008},
  \emph{129}\relax
\mciteBstWouldAddEndPuncttrue
\mciteSetBstMidEndSepPunct{\mcitedefaultmidpunct}
{\mcitedefaultendpunct}{\mcitedefaultseppunct}\relax
\EndOfBibitem
\bibitem[Myung \latin{et~al.}(2018)Myung, Roosen-Runge, Winkler, Gompper,
  Schurtenberger, and Stradner]{Myung2018}
Myung,~J.~S.; Roosen-Runge,~F.; Winkler,~R.~G.; Gompper,~G.;
  Schurtenberger,~P.; Stradner,~A. {Weak Shape Anisotropy Leads to a
  Nonmonotonic Contribution to Crowding, Impacting Protein Dynamics under
  Physiologically Relevant Conditions}. \emph{The Journal of Physical Chemistry
  B} \textbf{2018}, \emph{122}, 12396--12402\relax
\mciteBstWouldAddEndPuncttrue
\mciteSetBstMidEndSepPunct{\mcitedefaultmidpunct}
{\mcitedefaultendpunct}{\mcitedefaultseppunct}\relax
\EndOfBibitem
\bibitem[Roberts \latin{et~al.}(2014)Roberts, Keeling, Tracka, Van Der~Walle,
  Uddin, Warwicker, and Curtis]{roberts2014role}
Roberts,~D.; Keeling,~R.; Tracka,~M.; Van Der~Walle,~C.; Uddin,~S.;
  Warwicker,~J.; Curtis,~R. The role of electrostatics in protein--protein
  interactions of a monoclonal antibody. \emph{Molecular pharmaceutics}
  \textbf{2014}, \emph{11}, 2475--2489\relax
\mciteBstWouldAddEndPuncttrue
\mciteSetBstMidEndSepPunct{\mcitedefaultmidpunct}
{\mcitedefaultendpunct}{\mcitedefaultseppunct}\relax
\EndOfBibitem
\bibitem[Calero-Rubio \latin{et~al.}(2018)Calero-Rubio, Ghosh, Saluja, and
  Roberts]{calero2018predicting}
Calero-Rubio,~C.; Ghosh,~R.; Saluja,~A.; Roberts,~C.~J. Predicting
  protein-protein interactions of concentrated antibody solutions using dilute
  solution data and coarse-grained molecular models. \emph{Journal of
  pharmaceutical sciences} \textbf{2018}, \emph{107}, 1269--1281\relax
\mciteBstWouldAddEndPuncttrue
\mciteSetBstMidEndSepPunct{\mcitedefaultmidpunct}
{\mcitedefaultendpunct}{\mcitedefaultseppunct}\relax
\EndOfBibitem
\bibitem[Sibanda \latin{et~al.}(2023)Sibanda, Shanmugam, and
  Curtis]{sibanda2023relationship}
Sibanda,~N.; Shanmugam,~R.~K.; Curtis,~R. The relationship between
  protein--protein interactions and liquid--liquid phase separation for
  monoclonal antibodies. \emph{Molecular Pharmaceutics} \textbf{2023},
  \emph{20}, 2662--2674\relax
\mciteBstWouldAddEndPuncttrue
\mciteSetBstMidEndSepPunct{\mcitedefaultmidpunct}
{\mcitedefaultendpunct}{\mcitedefaultseppunct}\relax
\EndOfBibitem
\bibitem[Chowdhury \latin{et~al.}(2023)Chowdhury, Manohar, Lanzaro, Kimball,
  Witek, Woldeyes, Majumdar, Qian, Xu, Gillilan, \latin{et~al.}
  others]{chowdhury2023characterizing}
Chowdhury,~A.~A.; Manohar,~N.; Lanzaro,~A.; Kimball,~W.~D.; Witek,~M.~A.;
  Woldeyes,~M.~A.; Majumdar,~R.; Qian,~K.~K.; Xu,~S.; Gillilan,~R.~E.,
  \latin{et~al.}  Characterizing protein--protein interactions and viscosity of
  a monoclonal antibody from low to high concentration using small-angle x-ray
  scattering and molecular dynamics simulations. \emph{Molecular Pharmaceutics}
  \textbf{2023}, \emph{20}, 5563--5578\relax
\mciteBstWouldAddEndPuncttrue
\mciteSetBstMidEndSepPunct{\mcitedefaultmidpunct}
{\mcitedefaultendpunct}{\mcitedefaultseppunct}\relax
\EndOfBibitem
\bibitem[Roberts \latin{et~al.}(2014)Roberts, Keeling, Tracka, van~der Walle,
  Uddin, Warwicker, and Curtis]{Roberts2014}
Roberts,~D.; Keeling,~R.; Tracka,~M.; van~der Walle,~C.~F.; Uddin,~S.;
  Warwicker,~J.; Curtis,~R. {The Role of Electrostatics in Protein-Protein
  Interactions of a Monoclonal Antibody}. \emph{Molecular Pharmaceutics}
  \textbf{2014}, \emph{11}, 2475--2489\relax
\mciteBstWouldAddEndPuncttrue
\mciteSetBstMidEndSepPunct{\mcitedefaultmidpunct}
{\mcitedefaultendpunct}{\mcitedefaultseppunct}\relax
\EndOfBibitem
\bibitem[Gulotta \latin{et~al.}(2024)Gulotta, Polimeni, Lenton, Starr,
  Stradner, Zaccarelli, and Schurtenberger]{gulotta2024}
Gulotta,~A.; Polimeni,~M.; Lenton,~S.; Starr,~C.~G.; Stradner,~A.;
  Zaccarelli,~E.; Schurtenberger,~P. Combining scattering experiments and
  colloid theory to characterize charge effects in concentrated antibody
  solutions. \emph{Molecular Pharmaceutics} \textbf{2024}, \emph{21},
  2250--2271\relax
\mciteBstWouldAddEndPuncttrue
\mciteSetBstMidEndSepPunct{\mcitedefaultmidpunct}
{\mcitedefaultendpunct}{\mcitedefaultseppunct}\relax
\EndOfBibitem
\bibitem[Polimeni \latin{et~al.}(2024)Polimeni, Zaccarelli, Gulotta, Lund,
  Stradner, and Schurtenberger]{Polimeni2024}
Polimeni,~M.; Zaccarelli,~E.; Gulotta,~A.; Lund,~M.; Stradner,~A.;
  Schurtenberger,~P. A multi-scale numerical approach to study monoclonal
  antibodies in solution. \emph{APL bioengineering} \textbf{2024},
  \emph{8}\relax
\mciteBstWouldAddEndPuncttrue
\mciteSetBstMidEndSepPunct{\mcitedefaultmidpunct}
{\mcitedefaultendpunct}{\mcitedefaultseppunct}\relax
\EndOfBibitem
\bibitem[Denton(2003)]{Denton2003}
Denton,~A.~R. {Counterion penetration and effective electrostatic interactions
  in solutions of polyelectrolyte stars and microgels}. \emph{Physical Review
  E} \textbf{2003}, \emph{67}, 011804\relax
\mciteBstWouldAddEndPuncttrue
\mciteSetBstMidEndSepPunct{\mcitedefaultmidpunct}
{\mcitedefaultendpunct}{\mcitedefaultseppunct}\relax
\EndOfBibitem
\bibitem[Mahapatra \latin{et~al.}(2022)Mahapatra, Polimeni, Gentiluomo,
  Roessner, Frieß, Peters, Streicher, Lund, and Harris]{Mahapatra2022}
Mahapatra,~S.; Polimeni,~M.; Gentiluomo,~L.; Roessner,~D.; Frieß,~W.;
  Peters,~G. H.~J.; Streicher,~W.~W.; Lund,~M.; Harris,~P. Self-Interactions of
  Two Monoclonal Antibodies: Small-Angle X-ray Scattering, Light Scattering,
  and Coarse-Grained Modeling. \emph{Molecular Pharmaceutics} \textbf{2022},
  \emph{19}, 508--519, PMID: 34939811\relax
\mciteBstWouldAddEndPuncttrue
\mciteSetBstMidEndSepPunct{\mcitedefaultmidpunct}
{\mcitedefaultendpunct}{\mcitedefaultseppunct}\relax
\EndOfBibitem
\bibitem[Kaieda \latin{et~al.}(2014)Kaieda, Lund, Plivelic, and
  Halle]{Kaieda2014WeakSO}
Kaieda,~S.; Lund,~M.; Plivelic,~T.~S.; Halle,~B. Weak self-interactions of
  globular proteins studied by small-angle X-ray scattering and structure-based
  modeling. \emph{The journal of physical chemistry. B} \textbf{2014},
  \emph{118 34}, 10111--9\relax
\mciteBstWouldAddEndPuncttrue
\mciteSetBstMidEndSepPunct{\mcitedefaultmidpunct}
{\mcitedefaultendpunct}{\mcitedefaultseppunct}\relax
\EndOfBibitem
\bibitem[Stenqvist \latin{et~al.}(2013)Stenqvist, Thuresson, Kurut, V{\'a}cha,
  and Lund]{Stenqvist2013FaunusA}
Stenqvist,~B.; Thuresson,~A.; Kurut,~A.; V{\'a}cha,~R.; Lund,~M. Faunus ? a
  flexible framework for Monte Carlo simulation. \emph{Molecular Simulation}
  \textbf{2013}, \emph{39}, 1233 -- 1239\relax
\mciteBstWouldAddEndPuncttrue
\mciteSetBstMidEndSepPunct{\mcitedefaultmidpunct}
{\mcitedefaultendpunct}{\mcitedefaultseppunct}\relax
\EndOfBibitem
\bibitem[Johnson \latin{et~al.}(1994)Johnson, Panagiotopoulos, and
  Gubbins]{Johnson1994ReactiveCM}
Johnson,~J.~K.; Panagiotopoulos,~A.~Z.; Gubbins,~K.~E. Reactive canonical Monte
  Carlo : a new simulation technique for reacting or associating fluids.
  \emph{Molecular Physics} \textbf{1994}, \emph{81}, 717--733\relax
\mciteBstWouldAddEndPuncttrue
\mciteSetBstMidEndSepPunct{\mcitedefaultmidpunct}
{\mcitedefaultendpunct}{\mcitedefaultseppunct}\relax
\EndOfBibitem
\bibitem[Thurlkill \latin{et~al.}(2006)Thurlkill, Grimsley, Scholtz, and
  Pace]{Thurlkill2006}
Thurlkill,~R.~L.; Grimsley,~G.~R.; Scholtz,~J.~M.; Pace,~C.~N. pK values of the
  ionizable groups of proteins. \emph{Protein Science} \textbf{2006},
  \emph{15}, 1214--1218\relax
\mciteBstWouldAddEndPuncttrue
\mciteSetBstMidEndSepPunct{\mcitedefaultmidpunct}
{\mcitedefaultendpunct}{\mcitedefaultseppunct}\relax
\EndOfBibitem
\bibitem[Likos \latin{et~al.}(1998)Likos, L{\"{o}}wen, Watzlawek, Abbas,
  Jucknischke, Allgaier, and Richter]{Likos1998}
Likos,~C.~N.; L{\"{o}}wen,~H.; Watzlawek,~M.; Abbas,~B.; Jucknischke,~O.;
  Allgaier,~J.; Richter,~D. {Star Polymers Viewed as Ultrasoft Colloidal
  Particles}. \emph{Physical Review Letters} \textbf{1998}, \emph{80},
  4450--4453\relax
\mciteBstWouldAddEndPuncttrue
\mciteSetBstMidEndSepPunct{\mcitedefaultmidpunct}
{\mcitedefaultendpunct}{\mcitedefaultseppunct}\relax
\EndOfBibitem
\bibitem[Bergman \latin{et~al.}(2018)Bergman, Gnan, Obiols-Rabasa, Meijer,
  Rovigatti, Zaccarelli, and Schurtenberger]{Bergman2018}
Bergman,~M.~J.; Gnan,~N.; Obiols-Rabasa,~M.; Meijer,~J.-M.; Rovigatti,~L.;
  Zaccarelli,~E.; Schurtenberger,~P. A new look at effective interactions
  between microgel particles. \emph{Nature Comm.} \textbf{2018}, \emph{9},
  5039\relax
\mciteBstWouldAddEndPuncttrue
\mciteSetBstMidEndSepPunct{\mcitedefaultmidpunct}
{\mcitedefaultendpunct}{\mcitedefaultseppunct}\relax
\EndOfBibitem
\bibitem[McQuarrie(2000)]{McQuarrie2000}
McQuarrie,~D. \emph{{Statistical Mechanics}}; University Science Books,
  2000\relax
\mciteBstWouldAddEndPuncttrue
\mciteSetBstMidEndSepPunct{\mcitedefaultmidpunct}
{\mcitedefaultendpunct}{\mcitedefaultseppunct}\relax
\EndOfBibitem
\bibitem[Carnahan and Starling(1969)Carnahan, and Starling]{Carnahan1969}
Carnahan,~N.~F.; Starling,~K.~E. {Equation of State for Nonattracting Rigid
  Spheres}. \emph{J. Chem. Phys.} \textbf{1969}, \emph{51}, 635--636\relax
\mciteBstWouldAddEndPuncttrue
\mciteSetBstMidEndSepPunct{\mcitedefaultmidpunct}
{\mcitedefaultendpunct}{\mcitedefaultseppunct}\relax
\EndOfBibitem
\bibitem[Goodstein(1975)]{Goodstein1975}
Goodstein,~D. \emph{{States of Matter}}; Prentice Hall: New York, 1975\relax
\mciteBstWouldAddEndPuncttrue
\mciteSetBstMidEndSepPunct{\mcitedefaultmidpunct}
{\mcitedefaultendpunct}{\mcitedefaultseppunct}\relax
\EndOfBibitem
\bibitem[Zerah and Hansen(1986)Zerah, and Hansen]{Zerah1986}
Zerah,~G.; Hansen,~J.-P. {Self-consistent integral equations for fluid pair
  distribution functions: Another attempt}. \emph{The Journal of Chemical
  Physics} \textbf{1986}, \emph{84}, 2336--2343\relax
\mciteBstWouldAddEndPuncttrue
\mciteSetBstMidEndSepPunct{\mcitedefaultmidpunct}
{\mcitedefaultendpunct}{\mcitedefaultseppunct}\relax
\EndOfBibitem
\bibitem[Cardinaux \latin{et~al.}(2007)Cardinaux, Stradner, Schurtenberger,
  Sciortino, and Zaccarelli]{Cardinaux2007}
Cardinaux,~F.; Stradner,~A.; Schurtenberger,~P.; Sciortino,~F.; Zaccarelli,~E.
  {Modeling equilibrium clusters in lysozyme solutions}. \emph{Europhysics
  Letters (EPL)} \textbf{2007}, \emph{77}, 48004\relax
\mciteBstWouldAddEndPuncttrue
\mciteSetBstMidEndSepPunct{\mcitedefaultmidpunct}
{\mcitedefaultendpunct}{\mcitedefaultseppunct}\relax
\EndOfBibitem
\bibitem[Cardinaux \latin{et~al.}(2011)Cardinaux, Zaccarelli, Stradner,
  Bucciarelli, Farago, Egelhaaf, Sciortino, and Schurtenberger]{Cardinaux2011}
Cardinaux,~F.; Zaccarelli,~E.; Stradner,~A.; Bucciarelli,~S.; Farago,~B.;
  Egelhaaf,~S.~U.; Sciortino,~F.; Schurtenberger,~P. {Cluster-driven dynamical
  arrest in concentrated lysozyme solutions}. \emph{Journal of Physical
  Chemistry B} \textbf{2011}, \emph{115}, 7227--7237\relax
\mciteBstWouldAddEndPuncttrue
\mciteSetBstMidEndSepPunct{\mcitedefaultmidpunct}
{\mcitedefaultendpunct}{\mcitedefaultseppunct}\relax
\EndOfBibitem
\bibitem[N{\"{a}}gele(1996)]{Naegele1996}
N{\"{a}}gele,~G. {On the dynamics and structure of charge-stabilized
  suspensions}. \emph{Phys. Rep.} \textbf{1996}, \emph{272}, 215 -- 372\relax
\mciteBstWouldAddEndPuncttrue
\mciteSetBstMidEndSepPunct{\mcitedefaultmidpunct}
{\mcitedefaultendpunct}{\mcitedefaultseppunct}\relax
\EndOfBibitem
\bibitem[Banchio and N{\"{a}}gele(2008)Banchio, and N{\"{a}}gele]{Banchio2008}
Banchio,~A.~J.; N{\"{a}}gele,~G. {Short-time transport properties in dense
  suspensions: From neutral to charge-stabilized colloidal spheres}. \emph{J.
  Chem. Phys.} \textbf{2008}, \emph{128}, 104903\relax
\mciteBstWouldAddEndPuncttrue
\mciteSetBstMidEndSepPunct{\mcitedefaultmidpunct}
{\mcitedefaultendpunct}{\mcitedefaultseppunct}\relax
\EndOfBibitem
\bibitem[Yearley \latin{et~al.}(2013)Yearley, Zarraga, Shire, Scherer, Gokarn,
  Wagner, and Liu]{Yearley2013}
Yearley,~E.~J.; Zarraga,~I.~E.; Shire,~S.~J.; Scherer,~T.~M.; Gokarn,~Y.;
  Wagner,~N.~J.; Liu,~Y. {Small-Angle Neutron Scattering Characterization of
  Monoclonal Antibody Conformations and Interactions at High Concentrations}.
  \emph{Biophys. J.} \textbf{2013}, \emph{105}, 720--731\relax
\mciteBstWouldAddEndPuncttrue
\mciteSetBstMidEndSepPunct{\mcitedefaultmidpunct}
{\mcitedefaultendpunct}{\mcitedefaultseppunct}\relax
\EndOfBibitem
\bibitem[Corbett \latin{et~al.}(2017)Corbett, Hebditch, Keeling, Ke, Ekizoglou,
  Sarangapani, Pathak, Walle, Uddin, Baldock, Avenda{\~{n}}o, and
  Curtis]{Corbett2017}
Corbett,~D.; Hebditch,~M.; Keeling,~R.; Ke,~P.; Ekizoglou,~S.; Sarangapani,~P.;
  Pathak,~J.; Walle,~C. F. V.~D.; Uddin,~S.; Baldock,~C.; Avenda{\~{n}}o,~C.;
  Curtis,~R.~A. {Coarse-Grained Modeling of Antibodies from Small-Angle
  Scattering Profiles}. \emph{J. Phys. Chem. B} \textbf{2017}, \emph{121},
  8276--8290\relax
\mciteBstWouldAddEndPuncttrue
\mciteSetBstMidEndSepPunct{\mcitedefaultmidpunct}
{\mcitedefaultendpunct}{\mcitedefaultseppunct}\relax
\EndOfBibitem
\bibitem[Pedersen(2001)]{Pedersen2001}
Pedersen,~J.~S. {Structure factors effects in small-angle scattering from block
  copolymer micelles and star polymers}. \emph{The Journal of Chemical Physics}
  \textbf{2001}, \emph{114}, 2839--2846\relax
\mciteBstWouldAddEndPuncttrue
\mciteSetBstMidEndSepPunct{\mcitedefaultmidpunct}
{\mcitedefaultendpunct}{\mcitedefaultseppunct}\relax
\EndOfBibitem
\bibitem[Corti and Degiorgio(1981)Corti, and Degiorgio]{Corti1981}
Corti,~M.; Degiorgio,~V. {Quasi-elastic light scattering study of intermicellar
  interactions in aqueous sodium dodecyl sulfate solutions}. \emph{The Journal
  of Physical Chemistry} \textbf{1981}, \emph{85}, 711--717\relax
\mciteBstWouldAddEndPuncttrue
\mciteSetBstMidEndSepPunct{\mcitedefaultmidpunct}
{\mcitedefaultendpunct}{\mcitedefaultseppunct}\relax
\EndOfBibitem
\bibitem[Lanzaro \latin{et~al.}(2021)Lanzaro, Roche, Sibanda, Corbett, Davis,
  Shah, Pathak, Uddin, Van Der~Walle, Yuan, Pluen, and Curtis]{Lanzaro2021}
Lanzaro,~A.; Roche,~A.; Sibanda,~N.; Corbett,~D.; Davis,~P.; Shah,~M.;
  Pathak,~J.~A.; Uddin,~S.; Van Der~Walle,~C.~F.; Yuan,~X.-F.; Pluen,~A.;
  Curtis,~R. Cluster {Percolation} {Causes} {Shear} {Thinning} {Behavior} in
  {Concentrated} {Solutions} of {Monoclonal} {Antibodies}. \emph{Molecular
  Pharmaceutics} \textbf{2021}, \emph{18}, 2669--2682\relax
\mciteBstWouldAddEndPuncttrue
\mciteSetBstMidEndSepPunct{\mcitedefaultmidpunct}
{\mcitedefaultendpunct}{\mcitedefaultseppunct}\relax
\EndOfBibitem
\bibitem[Hung \latin{et~al.}(2019)Hung, Dear, Karouta, Chowdhury, Godfrin,
  Bollinger, Nieto, Wilks, Shay, Ramachandran, Sharma, Cheung, Truskett, and
  Johnston]{Hung2019a}
Hung,~J.~J.; Dear,~B.~J.; Karouta,~C.~A.; Chowdhury,~A.~A.; Godfrin,~P.~D.;
  Bollinger,~J.~A.; Nieto,~M.~P.; Wilks,~L.~R.; Shay,~T.~Y.; Ramachandran,~K.;
  Sharma,~A.; Cheung,~J.~K.; Truskett,~T.~M.; Johnston,~K.~P. Protein-Protein
  Interactions of Highly Concentrated Monoclonal Antibody Solutions via Static
  Light Scattering and Influence on the Viscosity. \emph{The journal of
  physical chemistry. B} \textbf{2019}, \relax
\mciteBstWouldAddEndPunctfalse
\mciteSetBstMidEndSepPunct{\mcitedefaultmidpunct}
{}{\mcitedefaultseppunct}\relax
\EndOfBibitem
\bibitem[Bratko \latin{et~al.}(2002)Bratko, Striolo, Wu, Blanch, and
  Prausnitz]{Bratko2002}
Bratko,~D.; Striolo,~A.; Wu,~J.~Z.; Blanch,~H.~W.; Prausnitz,~J.~M.
  {Orientation-Averaged Pair Potentials between Dipolar Proteins or Colloids}.
  \emph{The Journal of Physical Chemistry B} \textbf{2002}, \emph{106},
  2714--2720\relax
\mciteBstWouldAddEndPuncttrue
\mciteSetBstMidEndSepPunct{\mcitedefaultmidpunct}
{\mcitedefaultendpunct}{\mcitedefaultseppunct}\relax
\EndOfBibitem
\end{mcitethebibliography}

\clearpage
\newpage
\begin{center}
\large
\textbf{Supporting Information: A soft penetrable sphere colloid model for the description of charge and excluded volume interactions in antibody solutions}

\normalsize
\bigskip
Peter Schurtenberger\textsuperscript{1,2}, Marco Polimeni\textsuperscript{1}, Sophia Marzouk\textsuperscript{3}, Robin Curtis\textsuperscript{3}, Emanuela Zaccarelli\textsuperscript{4, 5}, Anna Stradner\textsuperscript{1, 2}, \\
\medskip
\small
\textit{%
\textsuperscript{1}Division of Physical Chemistry, Department of Chemistry, Lund University, Lund, Sweden\\
\textsuperscript{2}LINXS Institute of Advanced Neutron and X-ray Science, Lund University, Lund, Sweden\\
\textsuperscript{3}Manchester Institute of Biotechnology, Department of Chemical Engineering, Faculty of Science and Engineering, The University of Manchester, Manchester M1 7DN, U.K.\\
\textsuperscript{4}Department of Physics, Sapienza University of Rome, Piazzale Aldo Moro 2, 00185 Roma, Italy\\
\textsuperscript{5}CNR Institute of Complex Systems, Uos Sapienza, Piazzale Aldo Moro 2, 00185 Roma, Italy\\
}

\end{center}

\normalsize
\bigskip
\renewcommand{\theequation}{S\arabic{equation}}\setcounter{equation}{0}
\renewcommand{\thefigure}{S\arabic{figure}}\setcounter{figure}{0}
\renewcommand{\thetable}{S\arabic{table}}\setcounter{table}{0}

\newpage
\subsection{Charge distribution from constant-pH one-protein simulations}

\begin{figure}[h!]
\centering
\begin{subfigure}{0.45\textwidth}
\centering
\includegraphics[width = \textwidth]{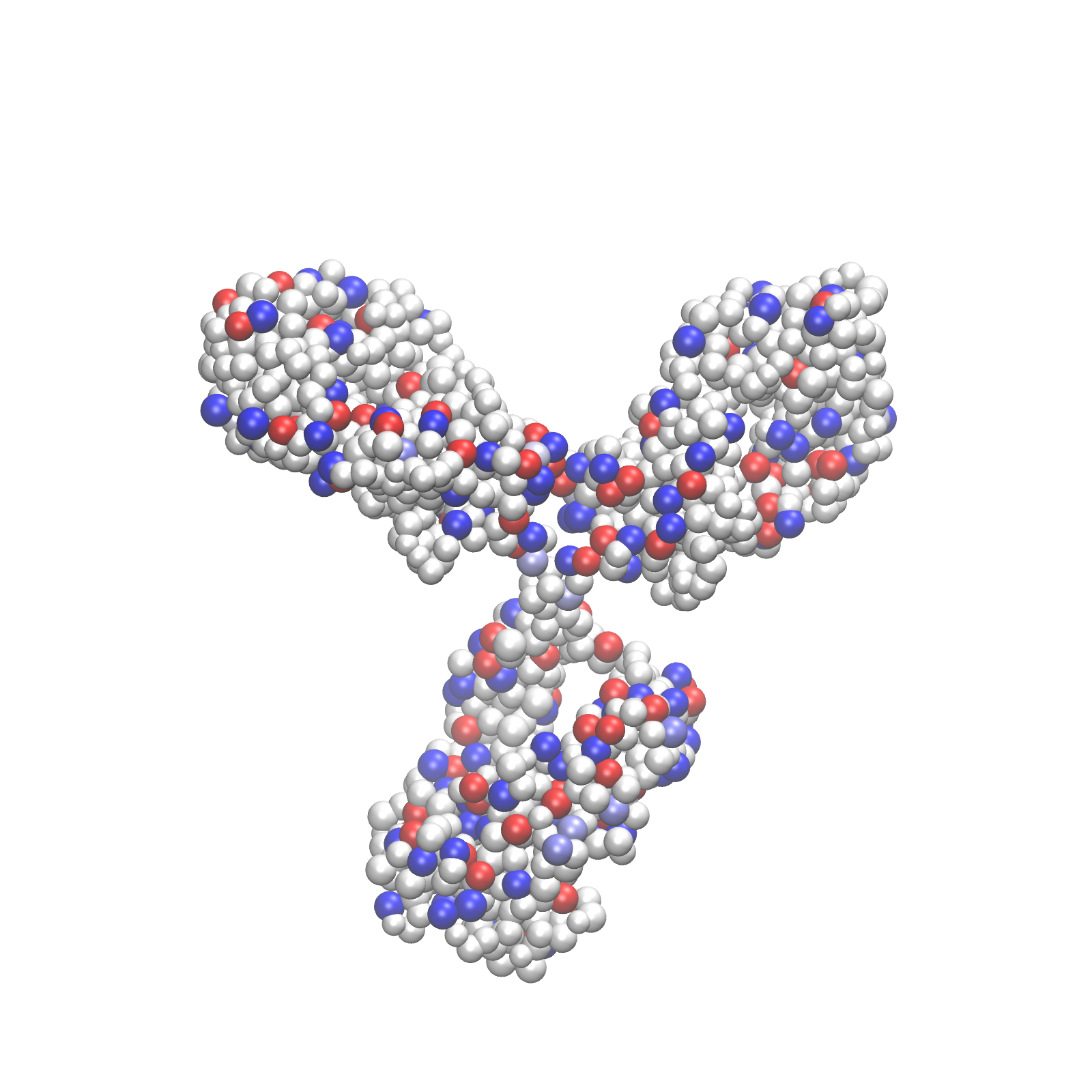}
\caption{}
\end{subfigure}
\begin{subfigure}{0.49\textwidth}
\centering
\includegraphics[width = \textwidth]{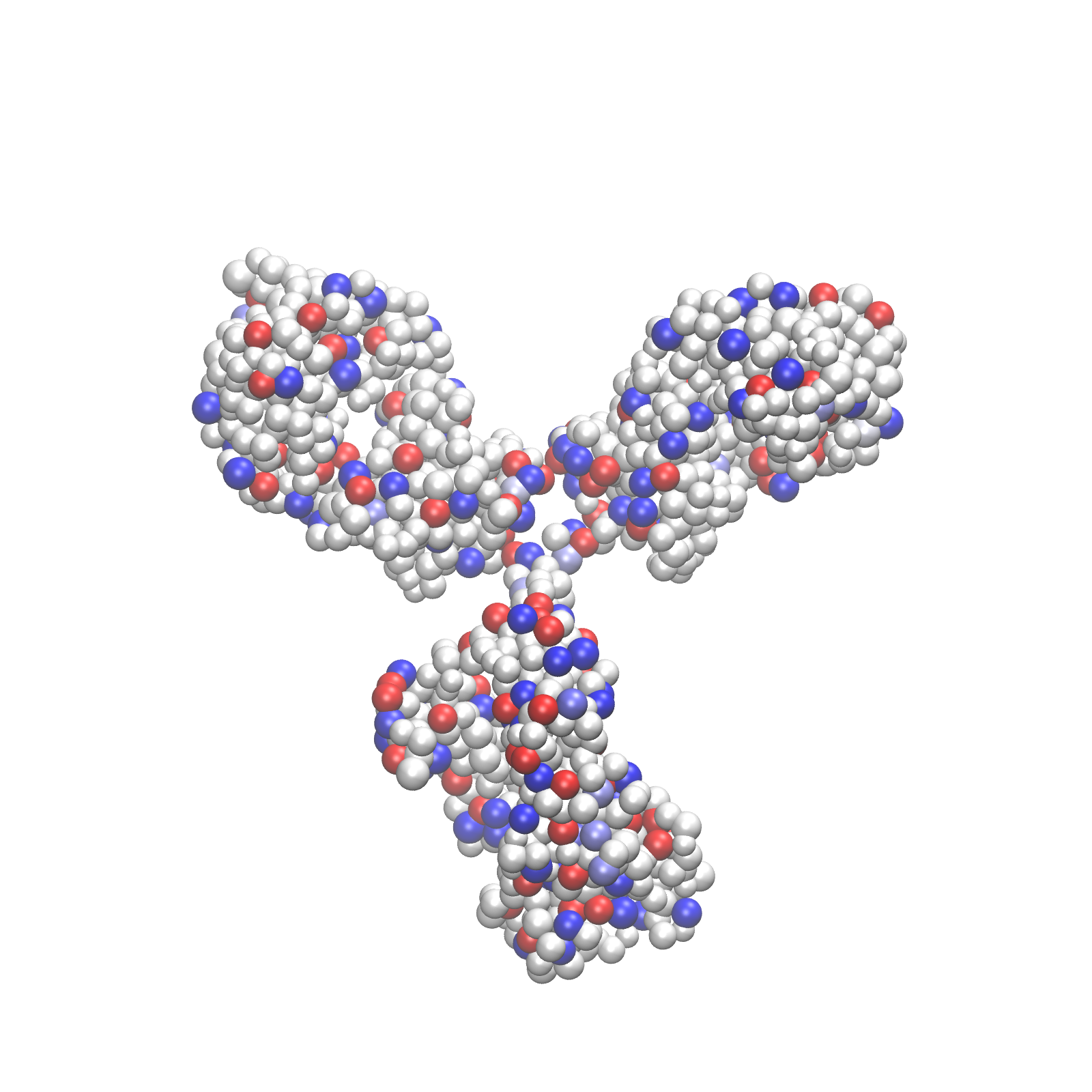}
\caption{}
\end{subfigure}
\caption{Charge distributions of mAb-1 at pH 6 and I=7 mM. Blue and red beads indicate, respectively, positively and negatively charged amino acids.}
\label{fig:combined1}
\end{figure}

\begin{figure}[h!]
\centering
\begin{subfigure}{0.45\textwidth}
\centering
\includegraphics[width = \textwidth]{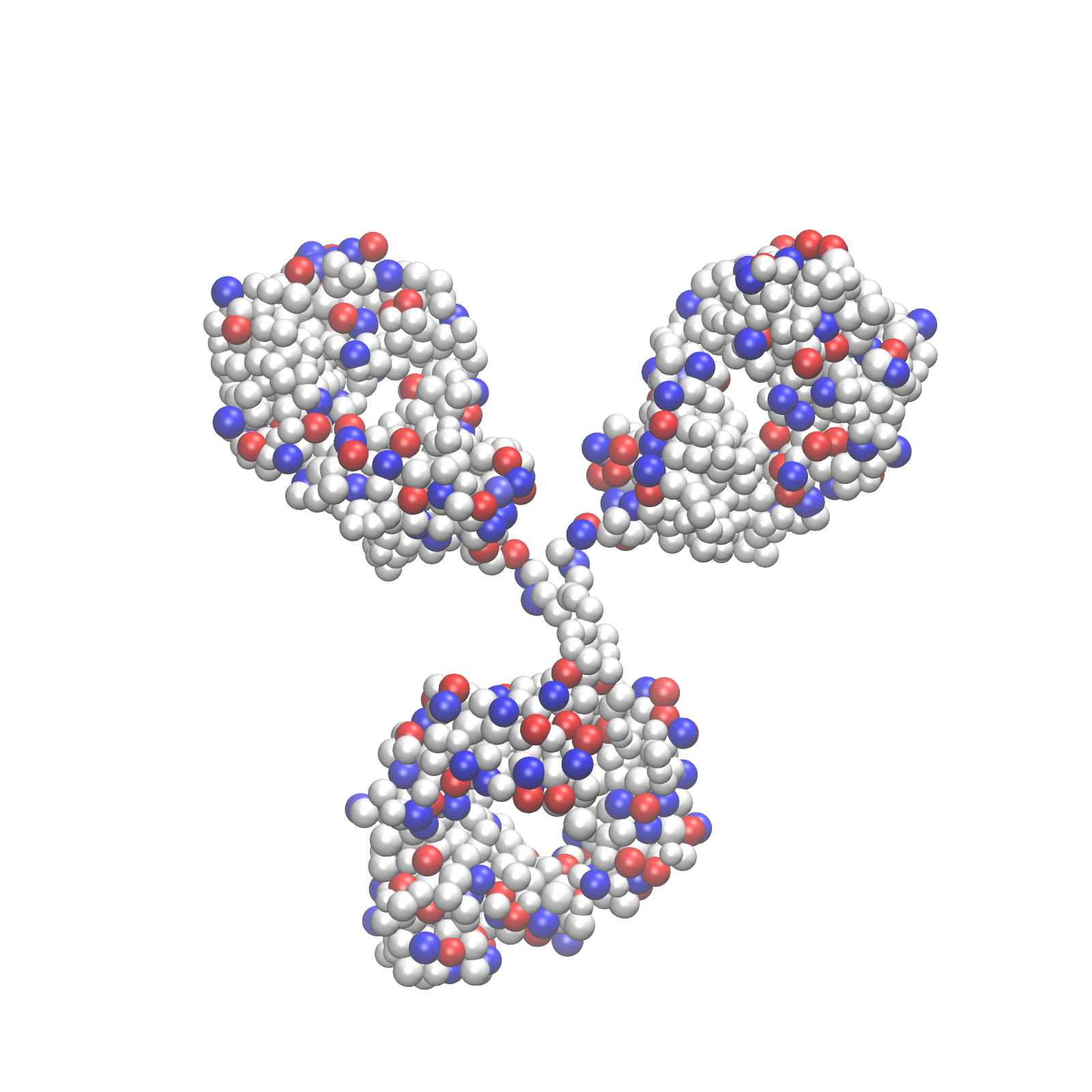}
\caption{}
\end{subfigure}
\begin{subfigure}{0.49\textwidth}
\centering
\includegraphics[width = \textwidth]{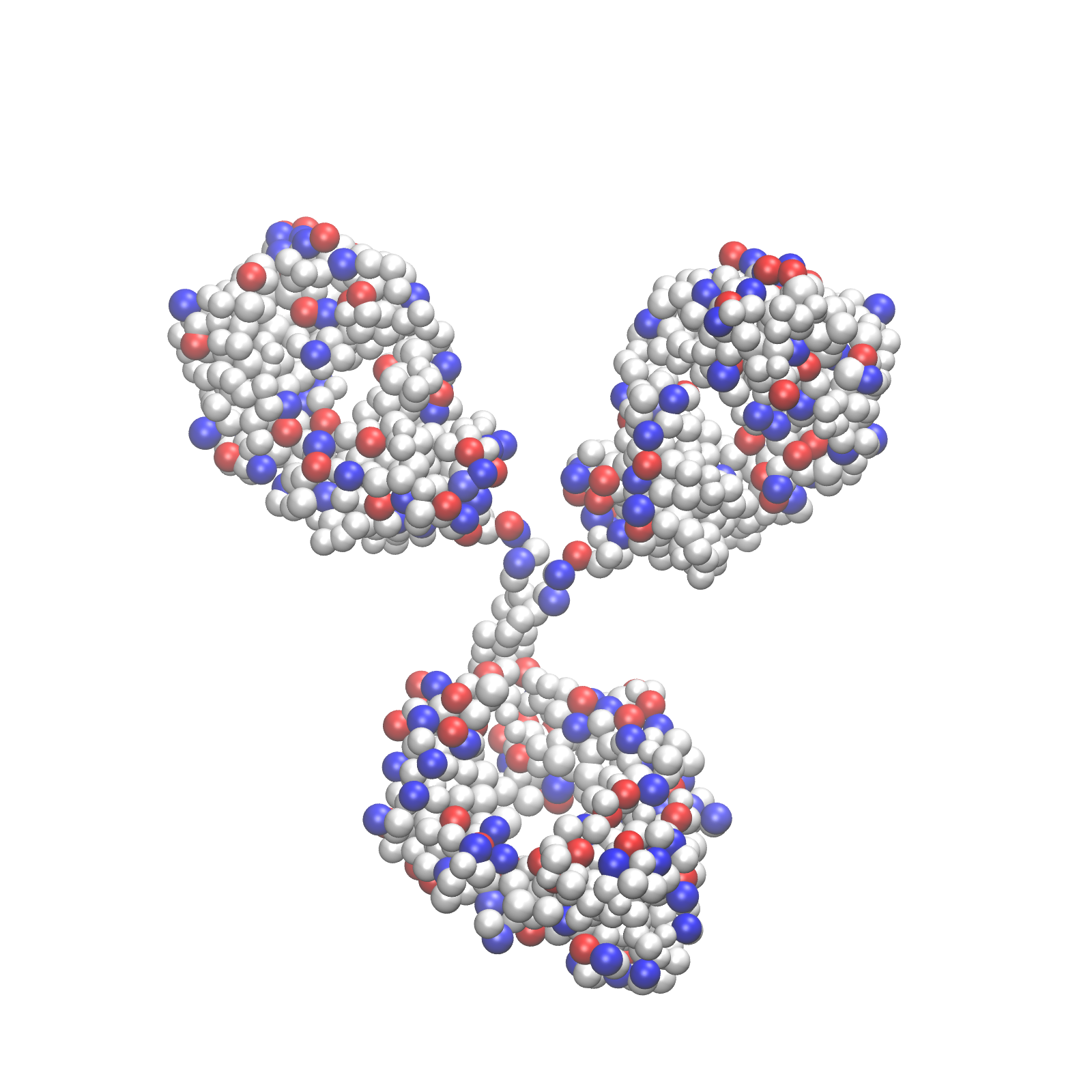}
\caption{}
\end{subfigure}
\caption{Charge distributions of mAb-2 at pH 5 and I=10 mM. Blue and red beads indicate, respectively, positively and negatively charged amino acids.}
\label{fig:combined2}
\end{figure}

\newpage

\subsection{Two-protein simulations - potential of mean force}

\begin{figure}[h!]
\centering
\includegraphics[width=0.8\linewidth]{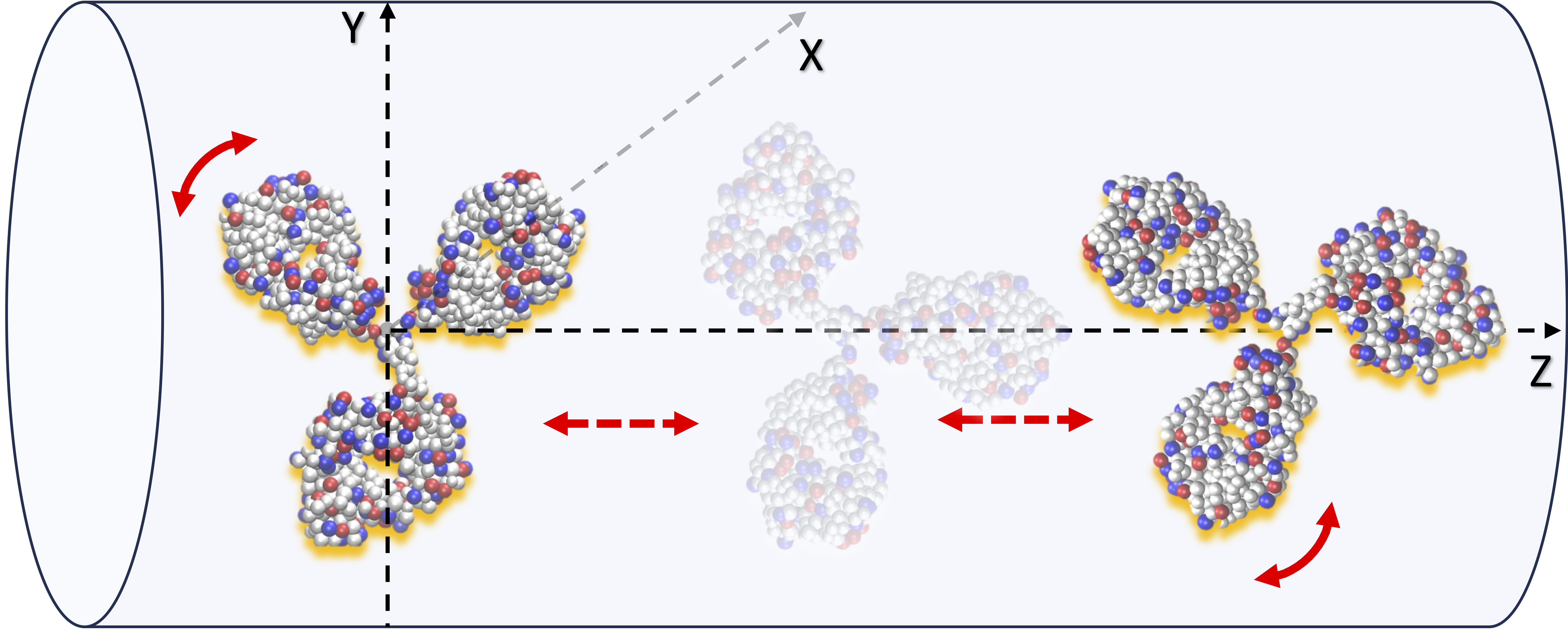}
\caption{Two-mAb simulations setup. }
\end{figure}

\subsection{Many-protein MC simulations}

\begin{figure}[h!]
\centering
\includegraphics[width=0.7\linewidth]{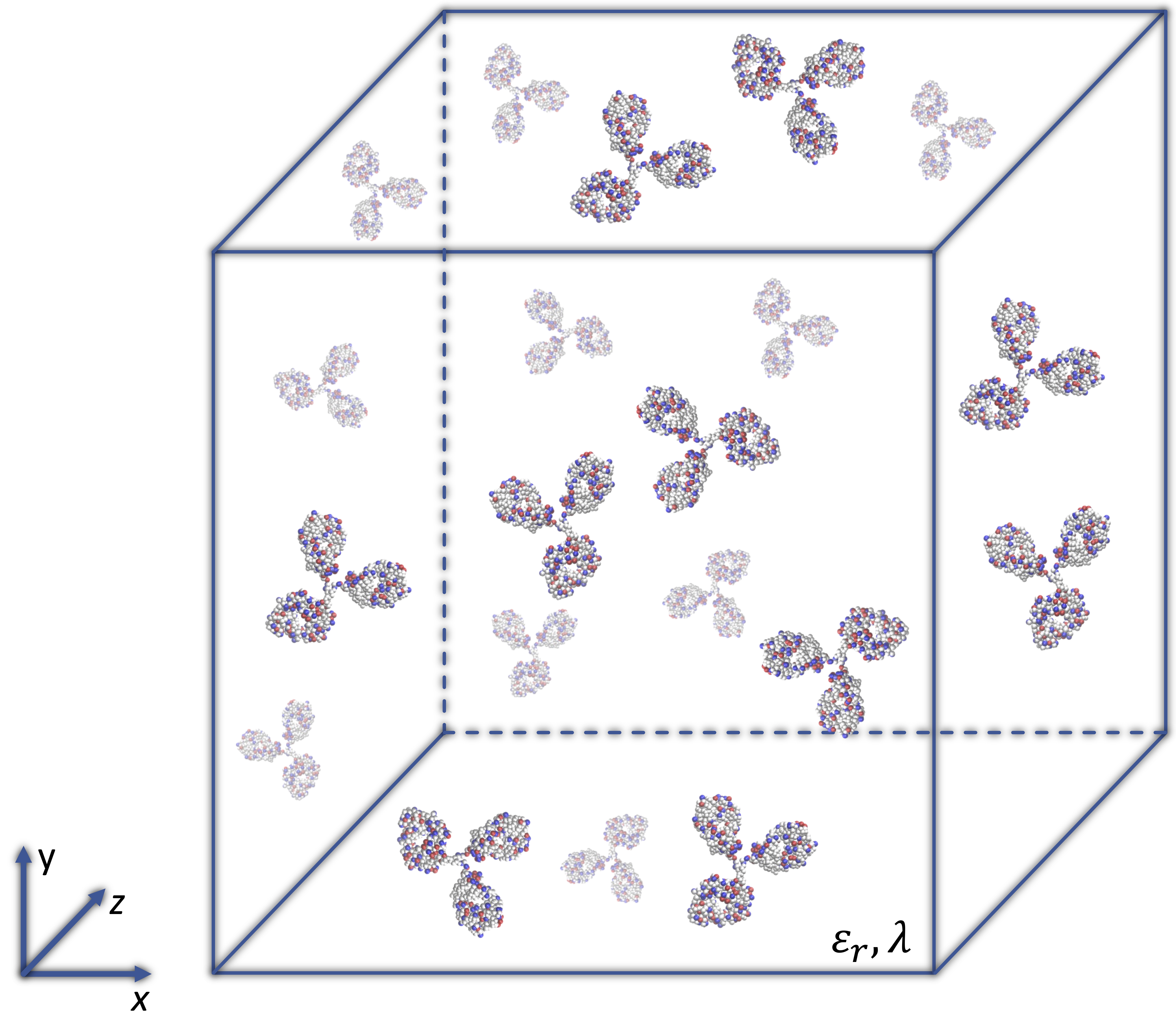}
\caption{Many-mAb simulations setup. }
\end{figure}

\subsection{Concentration dependence of the potential of mean force}

\begin{figure}[h]
\includegraphics[width=0.9\linewidth]{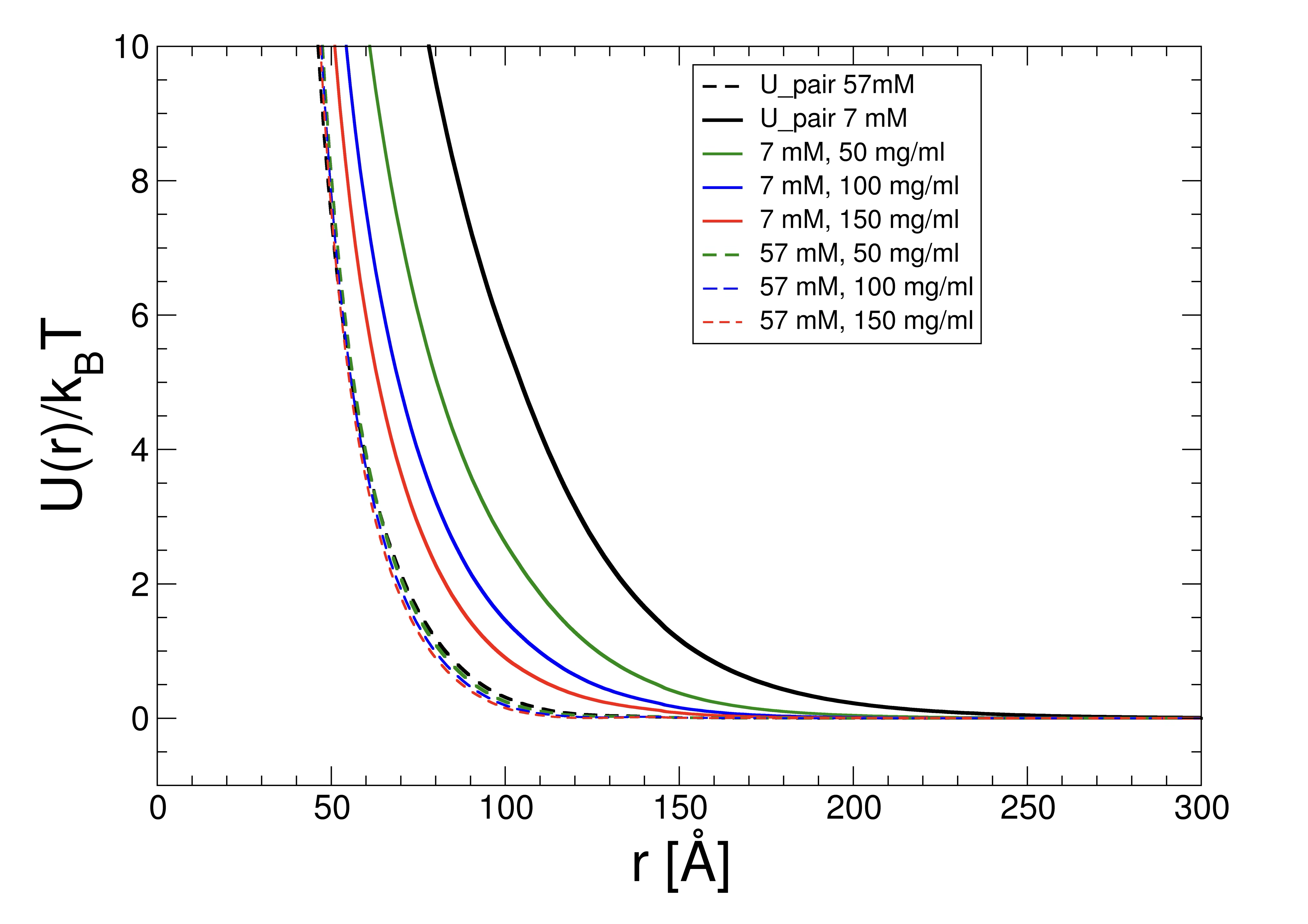}
\caption{Effective pair potential $U(r)/k_B T$ as a function of the centre-centre distance $r$ for different mAb concentrations and two ionic strengths (7 mM, 57 mM) for mAb-1: $U(r)$ in the limit of low concentrations (7 mM, solid black line; 57 mM, dashed black line), $c = 50$ mg/ml (7 mM, solid green line; 57 mM, dashed green line), $c = 100$ mg/ml (7 mM, solid blue line; 57 mM, dashed blue line), $c = 150$ mg/ml (7 mM, solid red line; 57 mM, dashed red line). See ref. \cite{gulotta2024} for details of the calculation of the effective ionic strength.}
\label{Potential}
\end{figure}

\newpage

\subsection{Concentration dependence of the effective charge $Z_{eff}$}

\begin{figure}[h]
\includegraphics[width=0.9\linewidth]{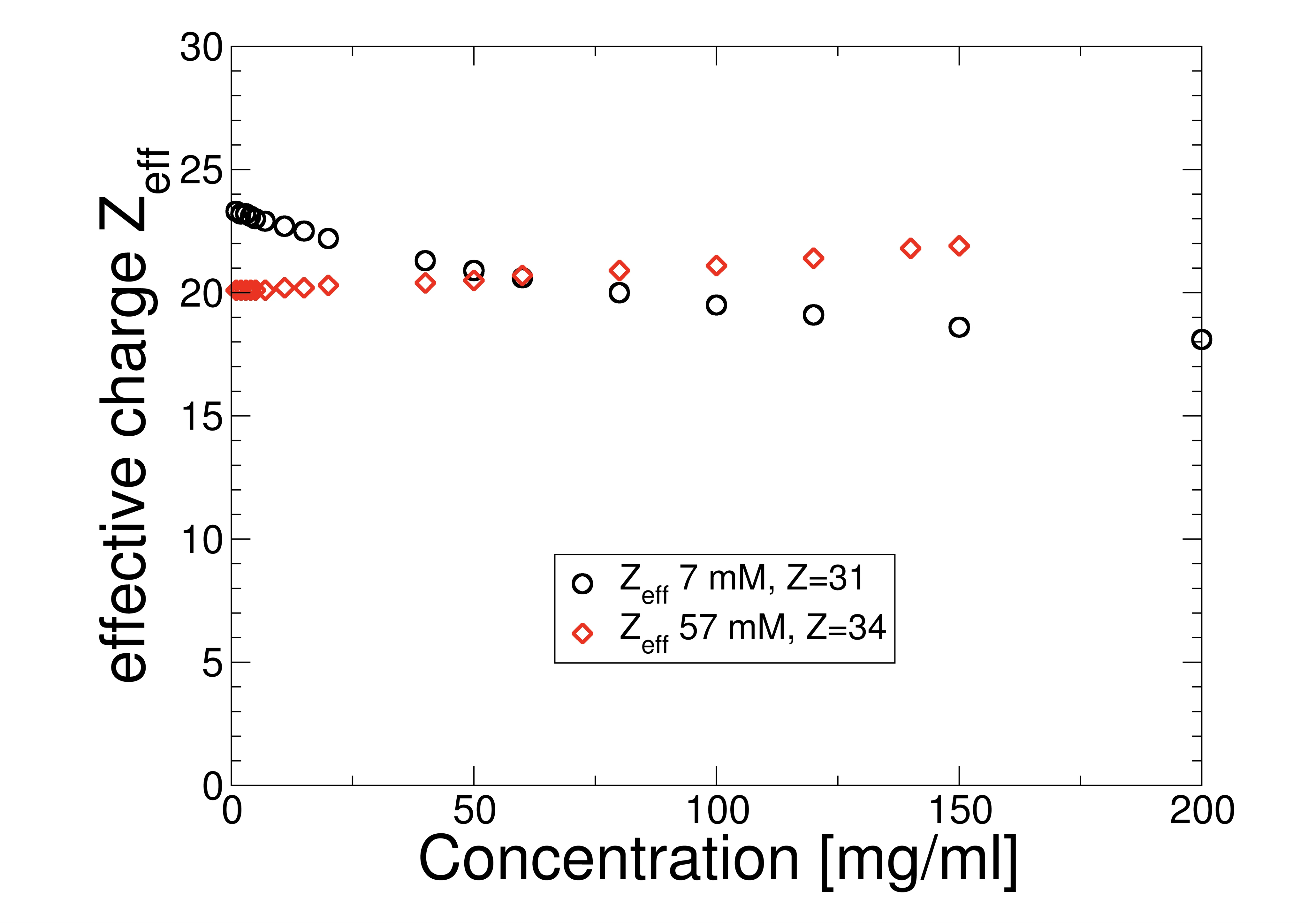}
\caption{Effective charge $Z_{eff}$ as a function of the mAb concentration for two ionic strengths (7 mM, open black circles; 57 mM, open red diamonds) for mAb-1. The effective charge $Z_{eff}$ is calculated for a total net charge $Z = 31$ at 7 mM and $Z = 34$ at 57 mM ionic strength using eqn. 15 in the main manuscript.}
\label{Zeff}
\end{figure}

\end{document}